\documentclass[12pt,onecolumn,draftclsnofoot]{IEEEtran}

\usepackage{color,array,amsthm}
\usepackage{graphicx}

\newtheorem{remark}{Remark}
\usepackage[utf8]{inputenc}
\usepackage{graphics} 
\usepackage{epsfig} 
\usepackage{mathptmx} 
\usepackage{times} 
\usepackage{amsmath} 
\usepackage{amssymb}  
\usepackage{booktabs}
\usepackage{float}
\usepackage{bm}
\usepackage{hyperref}
\usepackage{float}
\usepackage{caption}
\usepackage{subcaption}
\usepackage{xcolor}
\usepackage{soul}
\usepackage{siunitx}

\usepackage{array}
\usepackage{multirow}
\usepackage{multicol}
\usepackage{cite}

\graphicspath{ {Figures/} }

\begin{document}

	\title{ \Large \textbf{Data-Driven Reduced-Order Aeroelastic Modeling of Highly Flexible Aircraft by Parametric Dynamic Mode Decomposition} }

%
%
%
%
%
%
%
%

\author{Tianyi He,
	Weihua Su
	\thanks{He is with  Department of Mechanical and Aerospace Engineering, Utah State University, Logan, Utah 84322-4130, USA. Email: \href{mailto:tianyi.he@usu.edu}{tianyi.he@usu.edu}. He is the corresponding author.}
	\thanks{Su is with Department of Aerospace Engineering and Mechanics, The University of Alabama, Tuscaloosa, Alabama 35487-0280, USA. Email: \href{mailto:suw@eng.ua.edu}{suw@eng.ua.edu}.}
	}

\markboth{Journal of \LaTeX\ Class Files,~Vol.~14, No.~8, August~2015}%
{He \MakeLowercase{\textit{et al.}}}

	\maketitle
	
	\vspace{-4em}
\begin{abstract}
	\normalsize
This paper presents a method of data-driven parametric Dynamic Mode Decomposition (p-DMD) to derive a linear parameter-varying reduced-order model (LPV-ROM) for the nonlinear aeroelasticity of highly flexible aircraft. It directly uses the data snapshots obtained at varying flight conditions, and encodes the physical understanding of the nonlinear model's polynomial dependency on flight conditions to produce a polynomial-dependent LPV-ROM. Therefore, this method can handle not only the equilibrium flight conditions but also the cases of continuously-varying flight conditions. In the numerical studies, a highly flexible cantilever wing and a slender vehicle built based on it are first studied with fixed angles of attack as the scheduling parameter. The comparisons between traditional linearization-based parametric modeling and the data-driven p-DMD modeling are performed to verify the modeling accuracy. The results demonstrate that the current p-DMD modeling method can capture the aeroelastic and flight dynamic responses of highly flexible aircraft in both time and frequency domains. In addition, the proposed p-DMD method is applied to the highly flexible aircraft in a perturbed longitudinal flight with varying angles of attack as the scheduling parameter. The nonlinear aeroelastic and flight dynamic data are compared with the simulation results of the data-driven p-DMD model. The comparison results demonstrate that it can accurately capture the non-equilibrium (or transient) aeroelastic and flight dynamic behaviors of such slender vehicles.
	\end{abstract}

\vspace{-1em}
	\begin{IEEEkeywords} \vspace{-1em}
		Data-Driven Modeling; Linear Parameter-Varying System; Dynamic Mode Decomposition; Highly Flexible Aircraft
	\end{IEEEkeywords}
	
\section{Introduction}
\label{sec: intro}
	
High-aspect-ratio slender wings have been demanded to design next-generation high-altitude, long-endurance aircraft due to the benefits of improved vehicle performance and efficiency. However, the geometrically nonlinear effect pertinent to the highly flexible wings must be adequately addressed in modeling the open-loop dynamic behaviors, analyzing their stability characteristics, and the control system development for such slender vehicles. Accurately predicting the coupled nonlinear aeroelasticity and flight dynamics of highly flexible aircraft is essential for their design, optimization, and control. However, the complex fluid-structure interactions (FSI) induced by the geometric nonlinearity, as well as the coupled nonlinear aeroelastic and flight dynamic behavior, pose significant challenges to the nonlinear aeroelastic modeling and active control of such vehicles.

One of the significant challenges is the nonlinear elastic deformation of highly flexible wings, which makes the FSI more complicated, even in the scenario of a simple wing geometry. Linearization is the most commonly used method to predict aircraft stability and other open-loop performance, and obtain control-oriented models~\cite{shoor1989aeroelastic}. However, linearization is conducted assuming small vibrations about an equilibrium flight condition. Therefore, it cannot well address the vehicle's dynamic behavior with arbitrary large wing deformations. Linear parameter varying (LPV) modeling and control have been demonstrated to be a powerful tool that addresses the nonlinear aeroelasticity of flexible wing~\cite{barker2000comparing,he2023gust}. An LPV model can be established by interpolating multiple linearized models. After that, an LPV controller can be designed to be parameter-dependent to smoothly schedule controller gains at varying flight conditions or vibration magnitude~\cite{he2019smooth}.

Another challenge is the development of reduced-order models (ROMs) to enable fast computations of optimization and control, usually at the expense of fidelity. The high-fidelity structural model can be analyzed and reduced to a lower-order model, while the accuracy is almost maintained by keeping the most significant modes. An eigenvalue problem is usually solved, and a set of orthogonal modes are selected as states of the reduced-order model. Model reductions are repeated at gridded flight conditions, and interpolations are conducted to capture parametric dependency to get a parametric model over a range of flight conditions.  It's quite often to encounter the cases at varying flight conditions: 1) System stability varies from stable to unstable, or varies from unstable to stable~\cite{he2023gust}; 2) A complex conjugate pair merge to two real-valued poles or two real-valued poles split to complex conjugate pair~\cite{fonzi2020data}. Therefore, the eigenvalues and eigenvectors need to be tracked and aligned at gridded flight conditions~\cite{al2017lpv} to ensure the consistency of states among the fixed models.

Recent years have seen a growing interest in data-driven modeling techniques for the aeroelastic analysis of flexible wings. There are many data-driven methods of aeroelastic modeling using neural networks~\cite{li2019deep}, Kriging~\cite{timme2011transonic}. These methods leverage a large amount of data sets obtained from experiments or numerical simulations, train a model to learn the input-output relationships without a physical model, and achieve accurate predictions of aeroelastic behaviors. These approaches are particularly well-suited for complex, nonlinear systems where analytical models may be insufficiently precise or difficult to establish. However, the computational complexity is usually not time-efficient. Besides, these models lack an in-depth understanding and interpretation of the physical system. Therefore, these models are unreliable for flight control design, which is safety critical. 

State-space model by the data-driven method is of particular interest from the control point of view. Identification methods or mode decomposition can be used to find the optimal reduced-order model that best approximates the system. Eigensystem realization algorithm (ERA)~\cite{juang1985eigensystem} was developed for modal parameter identification and model reduction from testing data. Autoregressive with exogenous input (ARX) model was derived to identify the input-output relationships in discrete-time domain~\cite{zhang2007reduced,zhang2009aeroservoelastic}. Note that neither ARX nor ERA encodes the information of the physical states. If the physical state information is available, a physically-interpretable state-space model can be obtained via several methods. Proper orthogonal decomposition (POD) aims at finding a proper low-dimensional subspace of spatially orthogonal modes that best approximate the data in the $l_{2}$-norm manner~\cite{lumley1967structure,berkooz1993proper}. Sparse identification of nonlinear dynamics (SINDy) can produce a physically interpretable nonlinear model based on a library of nonlinear basis functions~\cite{brunton2016discovering,kaiser2018sparse}. 

Dynamic Mode Decomposition (DMD) is an equation-free method that can reveal the evolution of the spatial modes and temporal dynamics from the data snapshots. The well-established DMD algorithm uses data at perturbed equilibrium points, analyzes the mode, and renders a Linear Time-Invariant (LTI) model~\cite{iannelli2021balanced,fonzi2020data}. In order to obtain ROMs, the projection-based model reduction method was often employed to keep the most significant modes~\cite{benner2015survey}. A dynamic mode decomposition with control (DMDc) was applied on a flexible wing at different operating points and interpolated into a parametric model over a range of operating regimes~\cite {fonzi2020data}. A balanced mode decomposition (BMD) was derived using an oblique projection, where the projection basis space was shared to achieve state consistency, and the testing projection was chosen as parameter-dependent to obtain parametric models~\cite{iannelli2021balanced}. However, these works were based on the assumption of small wing deformations, and the persistently excited inputs were used to perturb the equilibrium points to obtain data snapshots. The transient system dynamics caused by the varying scheduling parameter were missing in the parametric model from interpolation. 

This paper presents an innovative method that directly uses data snapshots at varying flight conditions to derive a parametric LPV-ROM. The modeling accuracy is verified by comparing it with the traditional linearization-interpolation approach and nonlinear full-order models on the test bed of a slender wing and a flexible aircraft in our previous work~\cite{he2023gust}. The results reveal that, at fixed flight conditions of equilibrium, where a steady state is achieved, these two methods render consistent responses in the frequency domain and time domain. Furthermore, this paper demonstrates that the proposed method is better than traditional DMD because it can also address the scenario of varying flight conditions that are not equilibrium, where the aircraft exhibit transient dynamic behaviors.

The novelties of the proposed method and the contributions of this paper are three-fold: 
\begin{enumerate}
\item The proposed p-DMD is applicable not only at the equilibrium points, where a steady state is achieved, to derive an LTI-ROM, but also at the non-equilibrium scenarios to derive a polynomial-dependent LPV-ROM. The assumption of small deformation is relaxed because no small perturbation about the equilibrium is needed. Therefore, the proposed method can address the coupled aeroelasticity and flight dynamics for the highly flexible wing with large deformations and unsteady dynamics. Besides, p-DMD can capture the transient system dynamics at varying flight conditions.

\item It's well known that traditional parametric modeling by modal alignment, order reduction, and interpolations is challenging for large-dimension systems, usually resulted from high-fidelity finite-element modeling. However, the proposed method avoids these steps and derives parametric models directly from data snapshots, which makes p-DMD promising to extract ROMs from high-fidelity data. 

\item The physical understanding of parameter dependency is encoded into the parametric Hankel matrices of the data-driven LPV representation (see Eqns.~\eqref{data_poly_ss} and \eqref{data_representation}). The parametric ROM is directly obtained over a range of scheduling parameters without interpolations of grid models. Last, the state consistency is inherent under varying flight conditions because the same projection eigenvectors are used (see Eqn.~\eqref{parametric_ROM_projection}).

\end{enumerate}

In the rest of this paper, the traditional method of linearization-interpolation approach of aeroelastic modeling is briefly reviewed in Section~\ref{sec_linearization}. The data-driven reduced-order LPV modeling is then derived and described in Section~\ref{sec_pDMD_LPVROM}. After that, a numerical study and results will be provided to demonstrate the effectiveness of the proposed method in Sections~\ref{sec_numericalstudy1} and \ref{sec_numericalstudy2}. At last, conclusions are made in Section~\ref{sec_conclusion}.

\section{Nonlinear and Linearized Equations of Motion}
\label{sec_linearization}

For control system development of highly flexible aircraft, one must follow a systematic approach based on a nonlinear aeroelastic formulation. On top of that, the authors have developed reduced-order LPV models for highly flexible aircraft. This section briefly reviews the fundamentals of nonlinear aeroelastic formulation and LPV modeling by linearization and interpolation at equilibrium points.

There have been numerous endeavors that examine the nonlinear aeroelasticity of highly flexible aircraft. While more refined models based on shell formulations may capture more details of deformations in morphing and geometrically complicated wings, a geometrically nonlinear beam theory is usually sufficient to establish an aeroelastic formulation for highly flexible aircraft. Palacios et al.~\cite{palacios2010structural} comprehensively discussed three types of beam formulations for the structural, aeroelastic, and flight dynamic analysis of highly flexible aircraft. By coupling a strain-based geometrically nonlinear beam formulation with the finite-state inflow theory \cite{peters1994finite, peters1995finiteI, peters1995finite}, Cesnik and his colleagues established the nonlinear aeroelastic and flight dynamic formulation, now known as UM\slash NAST \cite{shearer2007nonlinear, su2010nonlinear, su2011dynamic}.

\begin{figure}[ht]
	\centering
	\includegraphics[width=0.45\linewidth]{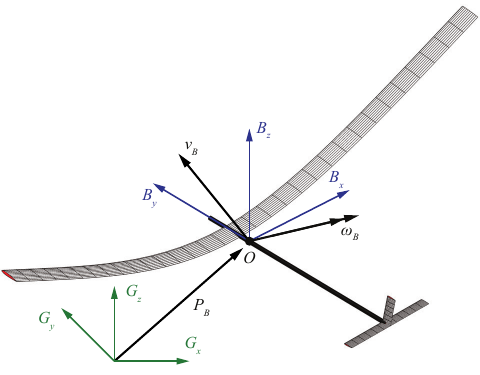}
	\caption{Flexible wing with large, nonlinear deformation.}
	\label{fig:101generichale01}
\end{figure}

By following the strain-based approach \cite{su2010nonlinear, su2011dynamic}, the coupled nonlinear aeroelasticity and flight dynamics of highly flexible aircraft (see Fig.~\ref{fig:101generichale01}) is governed by
\begin{equation} \label{eq: nl_ae_fd_eom}
	\begin{gathered}
		\mathbf{M}_{FF}(\bm{\varepsilon}) \ddot{\bm{\varepsilon}} + \mathbf{M}_{FB}(\bm{\varepsilon}) \dot{\bm{\beta}} + \mathbf{K}_{FF} \bm{\varepsilon} = \mathbf{R}_F \\ 
		\mathbf{M}_{BF}(\bm{\varepsilon}) \ddot{\bm{\varepsilon}} + \mathbf{M}_{BB}(\bm{\varepsilon}) \dot{\bm{\beta}} = \mathbf{R}_B \\ 
		\dot{\bm{\zeta}} = - \frac{1}{2} \bm{\Omega}_{\bm{\zeta}}(\bm{\beta}) \bm{\zeta} \\ 
		\dot{\mathbf{p}}_B^G = 
		\begin{bmatrix}
			\mathbf{C}^{GB}(\bm{\zeta}) & \mathbf{0}
		\end{bmatrix} 
		\bm{\beta} \\ 
		\dot{\bm{\mathbf{\lambda}}} = \mathbf{F}_1(\dot{\bm{\varepsilon}}, \bm{\beta}) \bm{\lambda} + 
		\mathbf{F}_2(\dot{\bm{\varepsilon}}, \bm{\beta})
		\begin{Bmatrix}
			\dot{\bm{\varepsilon}} \\ \bm{\beta} \\ 
		\end{Bmatrix} + 
		\mathbf{F}_3 
		\begin{Bmatrix}
			\ddot{\bm{\varepsilon}} \\ \dot{\bm{\beta}} \\ 
		\end{Bmatrix} \\ 
	\end{gathered}
\end{equation}
where the independent variables are the elastic strain along the flexible members $\bm{\varepsilon}$ and the aircraft rigid-body velocity $\bm{\beta}$. The rigid body's propagation is further described by the quaternion $\bm{\zeta}$ and inertial position $\mathbf{p}_B^G$. $\bm{\lambda}$ is the inflow state \cite{peters1994finite, peters1995finiteI, peters1995finite} from the unsteady aerodynamics. Finally, $\bm{\delta}$ consists of the deflections of trailing-edge control surfaces: elevator, aileron, and rudder, respectively. The load vector consists of the contributions of initial strain, damping, gravity, aerodynamics, and propulsion (thrust force), i.e.,
	\begin{equation} \label{eq: gen_loads}
		\begin{array}{c}
			\begin{Bmatrix}
			\mathbf{R}_F \\ 
			\mathbf{R}_B \\ 
		\end{Bmatrix} =
		\begin{Bmatrix}
			\mathbf{K}_{FF} \bm{\varepsilon}^0 \\ \mathbf{0} \\ 
		\end{Bmatrix} + 
		\begin{Bmatrix}
			\mathbf{R}_F^{\textrm{damp}}(\bm{\varepsilon}, \dot{\bm{\varepsilon}}, \bm{\beta}) \\ \mathbf{R}_B^{\textrm{damp}}(\bm{\varepsilon}, \dot{\bm{\varepsilon}}, \bm{\beta}) \\ 
		\end{Bmatrix} + 	\begin{Bmatrix}
		\mathbf{R}_F^{\textrm{grav}}(\bm{\zeta}) \\ \mathbf{R}_B^{\textrm{grav}}(\bm{\zeta}) \\ 
	\end{Bmatrix} \\
	+	\begin{Bmatrix}
			\mathbf{R}_F^{\textrm{aero}}(\bm{\varepsilon}, \dot{\bm{\varepsilon}}, \ddot{\bm{\varepsilon}}, \bm{\beta}, \dot{\bm{\beta}}, \bm{\lambda}, \bm{\delta}) \\ \mathbf{R}_B^{\textrm{aero}}(\bm{\varepsilon}, \dot{\bm{\varepsilon}}, \ddot{\bm{\varepsilon}}, \bm{\beta}, \dot{\bm{\beta}}, \bm{\lambda}, \bm{\delta}) \\
		\end{Bmatrix} +  
		\begin{Bmatrix}
			\mathbf{R}_F^{\textrm{prop}}(T) \\ \mathbf{R}_B^{\textrm{prop}}(T) \\ 
		\end{Bmatrix} \,.
		\end{array}
	\end{equation}
The detailed expressions are elaborated in \cite{su2010nonlinear, su2011dynamic, he2023gust}.

The traditional linearization-interpolation approach to derive an LPV model is summarized here. First, the nonlinear dynamical equation is linearized at a fixed equilibrium flight condition \cite{su2010nonlinear, he2023gust}. The second step is to perform the modal coordinate transformation on the high-dimensional full-order models (FOMs) at grid scheduling parameters. As a result, a series of LTI-FOMs in modal coordinates are obtained. The third step is to align the vibration modes to depict the modal evolution so that model reduction can be performed by keeping the most significant modes in terms of $H_{2}$ or $H_{\infty}$ norm. In the final step, the evolution of modal LTI-ROMs is interpolated into a polynomial parameter-dependent LPV-ROM \cite{al2017lpv}. An LPV controller and smooth switching LPV controller can be further developed against gust disturbance and model uncertainty \cite{he2019smooth}. 

The parametric state-space equation reads as 
	\begin{equation} \label{eq: linz_ss_sys_out}
		\begin{aligned}
			\dot{\mathbf{x}} & = \mathbf{A}(\theta) \mathbf{x} + \mathbf{B}(\theta) \mathbf{u} \hfill \\ 
			\mathbf{y} & = \mathbf{C}(\theta) \mathbf{x} + \mathbf{D}(\theta) \mathbf{u} \hfill \\ 
		\end{aligned}
	\end{equation}
	where the state variable and control input are
	\begin{equation} \label{eq: linz_ss_sys_x_u}
		\begin{aligned}
			\mathbf{x}^T & = \begin{Bmatrix}
				\Delta \bm{\varepsilon}^T & \Delta \dot{\bm{\varepsilon}}^T & \Delta \bm{\beta}^T & \Delta \bm{\zeta}^T & \left( \Delta \mathbf{p}_B^G \right)^T & \Delta \bm{\lambda}^T \\ 
			\end{Bmatrix} \hfill \\ 
			\mathbf{u}^T & = \begin{Bmatrix}
				\Delta \bm{\delta}^T & \Delta T \\ 
			\end{Bmatrix} \,. \hfill \\ 
		\end{aligned}
	\end{equation}
The scheduling parameter $\theta$ is usually a measurable quantity relevant to the varying flight condition, and output matrices $\mathbf{C}, \mathbf{D}$ are selected according to the user's need. The discrete-time model can be further converted from continuous-time model \eqref{eq: linz_ss_sys_out}.

This approach realizes the modeling of interactions among rigid-body dynamics, structural dynamics, and aerodynamics. The coupling affects the equivalent stiffness and inertial matrices under varying flight conditions. The model-based approach has several characteristics. First, this approach produces physically explainable models where the states of strains, rigid-body dynamics, and aerodynamics are aligned and analyzed. The state consistency across flight regimes is guaranteed. Second, the local stability of the coupled system can be easily analyzed by the linearized system matrices.


However, there are several limitations in the linearization-based approach: 1) The linearization and model reduction process usually need heavy computations if refined meshing is conducted at structures and aerodynamics nodes. Modal alignment to enable state consistency is difficult for high-dimensional systems, which is especially critical in high-fidelity modeling and unsteady aerodynamics. 2) The partial derivative of aerodynamic loading ($\mathbf{R}_F^{\textrm{aero}}$) relative to strains, strain rates need to be known. In the low-fidelity case, a complex analytical function can lead to close-form partial derivatives. However, the derivatives are hard to derive in high-fidelity and complex-geometry cases. 3) The biggest drawback of the linearization-based approach is that the deformation is assumed to be small around equilibrium. With geometrically nonlinear deformation or at transient flight conditions, the linearization approach cannot be applied. The LPV model from interpolation neglects the variation rate of the scheduling parameter and may cause pitfalls of the gain-scheduling controller.  

Therefore, in this study, we propose an innovative data-driven modeling approach for highly flexible aircraft that is capable of describing the geometrically nonlinear deformation in a dynamic flight condition. The input of the proposed algorithm is the data snapshots of states, scheduling parameters (flight conditions), and control inputs of persistent excitation. The outcome of the algorithm is a polynomial-dependent LPV-ROM. Users can select the dimension of the LPV-ROM and the polynomial order to describe the nonlinear aeroelasticity based on the need for accuracy levels. The data snapshots are not required to be collected at equilibrium points, but can also be collected at varying flight conditions. 

\section{Reduced-Order LPV Modeling by p-DMD}\label{sec_pDMD_LPVROM}
\subsection{Problem Statement}

Based on Eq.~\eqref{eq: nl_ae_fd_eom}, one can form a generic discrete-time nonlinear parameter varying model Eq.~\eqref{NPV_DT} to describe the flight dynamic and aeroelastic behavior of a highly flexible  aircraft.

\begin{equation}\label{NPV_DT}
	\begin{aligned}
		\mathbf{x}_{k+1} & =\mathbf{f}\left(\mathbf{x}_k, \mathbf{u}_k, \mathbf{\theta}_k\right) \\
		\mathbf{y}_k & = \mathbf{C}\mathbf{x}_{k}
	\end{aligned}
\end{equation}
where $\mathbf{x}_{k} \in \mathbf{R}^{n_{x}}$, $\mathbf{u}_{k} \in \mathbf{R}^{n_{u}}$, $\mathbf{y}_{k} \in \mathbf{R}^{n_{y}}$ are the states, control inputs, and outputs, with $k$ being the time index. $\mathbf{\theta}_{k}$ is a vector of time-varying scheduling parameters of flight conditions, which are assumed to be known in real-time. The output matrix $\mathbf{C}$ can be properly selected to obtain some of the states as outputs, for example, strains or rigid-body velocities, which renders a constant $\mathbf{C}$ and $\mathbf{D} = 0$.

The aircraft's aeroelastic behavior may vary with flight conditions, such as flight speed and angle of attack. This study considers only a single parameter, but it can be easily extended to a multiple-dimensional case. From the nonlinear model~\eqref{eq: nl_ae_fd_eom}, all states (strains, strain rates, quaternions, rigid-body motions, aerodynamic states) are included in the parametric FOM. Over a range of flight conditions $\theta_{k} \in \mathcal{P} \in \mathbf{R}$, the full-order nonlinear model~\eqref{NPV_DT} is approximated by a parametric FOM \eqref{FOM_ss}, and the matrices $\mathbf{A}(\mathbf{\theta}_{k}) \in \mathbf{R}^{n_{x} \times n_{x}} , \mathbf{B}(\mathbf{\theta}_{k}) \in \mathbf{R}^{n_{x} \times n_{u}}$ are assumed as $n_{p}$-order polynomials dependent on $\theta_{k}$ in~\eqref{poly_depen}. 

\begin{equation}\label{FOM_ss}
	\begin{aligned}
		\mathbf{x}_{k+1} & =\mathbf{A}\left(\mathbf{\theta}_k\right) \mathbf{x}_k+\mathbf{B}\left(\mathbf{\theta}_k\right) \mathbf{u}_k \\
		\mathbf{y}_k & =\mathbf{C} \mathbf{x}_k
	\end{aligned}
\end{equation}

\begin{equation}\label{poly_depen}
	\mathbf{A}\left(\mathbf{\theta}_k\right)=\mathbf{A}_0+\sum_{i=1}^{n_{\mathrm{p}}} (\theta_k)^{i} \mathbf{A}_i, \quad \mathbf{B}\left(\mathbf{\theta}_k\right)=\mathbf{B}_0+\sum_{i=1}^{n_{\mathrm{p}}} (\mathbf{\theta}_k)^{i} \mathbf{B}_i. 
\end{equation}

The technical problem is to find a reduced-order LPV state-space model~\eqref{ROM_ss} to approximate the nonlinear aeroelastic model, where matrices $\mathbf{A}_{r}(\theta_{k}) \in \mathbf{R}^{n_{z} \times n_{z}} , \mathbf{B}_{r}(\theta_{k}) \in \mathbf{R}^{n_{z} \times n_{u}}$, $n_{z} \ll n_{x}$, and $\mathbf{z}_{k}$ is the state vector of LPV-ROM.  

\begin{equation}\label{ROM_ss}
	\begin{aligned}
		\mathbf{z}_{k+1} & =\mathbf{A}_{r}\left(\theta_k\right) \mathbf{z}_k+\mathbf{B}_{r}\left(\theta_k\right) \mathbf{u}_k \\
		\mathbf{y}_k & =\mathbf{C}_{r} \mathbf{z}_k
	\end{aligned}
\end{equation}

\begin{equation}\label{poly_depen_ROM}
	\mathbf{A}_{r}\left(\theta_k\right)=\mathbf{A}_{r0}+\sum_{i=1}^{n_{\mathrm{p}}} (\theta_k)^{i} \mathbf{A}_{ri}, \quad \mathbf{B}_{r}\left(\theta_k\right)=\mathbf{B}_{r0}+\sum_{i=1}^{n_{\mathrm{p}}} (\theta_k)^{i} \mathbf{B}_{ri}. 
\end{equation}

\subsection{Parametric Data-Driven Representations }
Suppose the data snapshots are obtained as $\left\{\mathbf{u}_{k}, \mathbf{x}_{k}, \theta_{k} \right\}_{N_{d}}$ during time steps $k = 1$ to $N_{d}$.  The parametric LPV model~\eqref{FOM_ss}, \eqref{poly_depen} can be equivalently expressed by~\eqref{data_poly_ss} with parametric Hankel matrices.

\begin{equation}\label{data_poly_ss}
	\begin{array}{rl}
		\mathbf{x}_{k+1} = & (\mathbf{A}_0+\sum_{i=1}^{n_{\mathrm{p}}} (\theta_k)^{i} \mathbf{A}_i)x_{k} + (\mathbf{B}_0+\sum_{i=1}^{n_{\mathrm{p}}} (\theta_k)^{i} \mathbf{B}_i)\mathbf{u}_{k}\\
		= & \underbrace{\left[\begin{array}{cccc}
				\mathbf{A}_0 & \mathbf{A}_{1} & \cdots & \mathbf{A}_{n_{\mathrm{p}}}
			\end{array}\right]}_{\mathcal{A}}\left[\begin{array}{c}
			\mathbf{x}_k \\
			(\theta_k)^{1} \otimes \mathbf{x}_k \\ \vdots \\ 		(\theta_k)^{n_\mathrm{p}} \otimes \mathbf{x}_k
		\end{array}\right] \\ 
	+ & \underbrace{\left[\begin{array}{cccc}
				\mathbf{B}_0 & \mathbf{B}_{1} & \cdots & \mathbf{B}_{n_{\mathrm{p}}}
			\end{array}\right]}_{\mathcal{B}}\left[\begin{array}{c}
			\mathbf{u}_k \\
			(\theta_k)^{1} \otimes \mathbf{u}_k \\ \vdots \\ 		(\theta_k)^{n_\mathrm{p}} \otimes \mathbf{u}_k
		\end{array}\right] \\
		= & \mathcal{A} \underbrace{\left[\begin{array}{c}
			\mathbf{x}_k \\
			\Theta_k \otimes \mathbf{x}_k
		\end{array}\right]}_{\Omega_{\mathbf{x}}(\mathbf{\theta}_{k})} + \mathcal{B}\underbrace{ \left[\begin{array}{c}
			\mathbf{u}_k \\
			\Theta_k \otimes \mathbf{u}_k
		\end{array}\right]}_{\Omega_{\mathbf{u}}(\mathbf{\theta}_{k})} \\ 
	\end{array}
\end{equation}
where, $\Theta_k = \left[ (\theta_{k})^{1}, \cdots,  (\theta_{k})^{n_\mathrm{p}}\right]^{T}$. 

\begin{remark}[parametric Hankel matrix]
	The parametric data-driven representation of a nonlinear parameter-varying system has additional polynomial coefficient matrices $\mathbf{A}_{1}, \cdots, \mathbf{A}_{n_{\mathrm{p}}}, \mathbf{B}_{1}, \cdots, \mathbf{B}_{n_{\mathrm{p}}}$ other than $\mathbf{A}_{0}, \mathbf{B}_{0}$. The data snapshot formulates parametric Hankel matrices $\Omega_{\mathbf{x}}(\mathbf{\theta}_{k})$, $\Omega_{\mathbf{u}}(\mathbf{\theta}_{k})$, which encodes the physical understanding of the nonlinear aeroelasticity by the Kronecker product kernel. 
\end{remark}

Note that the dimension of $\Theta_{k}$ is $n_{p}$, and the dimension of $\mathbf{x}_{k}$ is $n_{x}$.  The following matrices can be constructed by stacking data snapshots of the sampling window from 1 to $N_{d}$.
\begin{equation}
	\begin{aligned}
		& \mathbf{U}=\left[\begin{array}{lll}
			\mathbf{u}_1, & \cdots, & \mathbf{u}_{N_{\mathrm{d}}}
		\end{array}\right], \mathbf{U}^{\mathrm{kr}}=\left[\Theta_1 \otimes \mathbf{u}_1, \cdots, \Theta_{N_{\mathrm{d}}} \otimes \mathbf{u}_{N_{\mathrm{d}}}\right], \\
		& \mathbf{X}=\left[\begin{array}{lll}
			\mathbf{x}_1, & \cdots, & \mathbf{x}_{N_{\mathrm{d}}}
		\end{array}\right], \mathbf{X}^{\mathrm{kr}}=\left[
		\Theta_1 \otimes \mathbf{x}_1, \cdots, \Theta_{N_{\mathrm{d}}} \otimes \mathbf{x}_{N_{\mathrm{d}}}
		\right], \\
		& \mathbf{X}^{+}=\left[\begin{array}{lll}
			\mathbf{x}_2 & \cdots & \mathbf{x}_{N_{\mathrm{d}}+1}
		\end{array}\right].
	\end{aligned}
\end{equation}

\begin{equation}
	\begin{array}{rl}
		\mathbf{X}^{+} & = \left[\begin{array}{ccc}
			\mathbf{x}_2 & \cdots & \mathbf{x}_{N_{\mathrm{d}}+1}
		\end{array}\right] \\
		& = \mathcal{A} \left[ \begin{array}{ccccc}
			\begin{array}{c}
				\mathbf{x}_1 \\
				\Theta_1 \otimes \mathbf{x}_1  
			\end{array} & \cdots \begin{array}{c}
				\mathbf{x}_{N_{d}} \\
				\Theta_{N_{d}} \otimes \mathbf{x}_{N_{d}} 
			\end{array}
		\end{array}\right]  + \mathcal{B} \left[ \begin{array}{ccc}
			\begin{array}{c}
				\mathbf{u}_1 \\
				\Theta_1 \otimes \mathbf{u}_1  
			\end{array} & \cdots \begin{array}{c}
				\mathbf{u}_{N_{d}} \\
				\Theta_{N_{d}} \otimes \mathbf{u}_{N_{d}} 
			\end{array}
		\end{array}\right].
	\end{array}
\end{equation}
Therefore, we have the full-order state evolution represented by the finite-horizon data snapshot, 
\begin{equation}
	\mathbf{X}^{+}=\mathcal{A}\left[\begin{array}{c}
		\mathbf{X} \\
		\mathbf{X}^{\mathrm{kr}}
	\end{array}\right]+\mathcal{B}\left[\begin{array}{c}
		\mathbf{U} \\
		\mathbf{U}^{\mathrm{kr}}
	\end{array}\right]  = \left[\mathcal{A} \quad \mathcal{B} \right] \left[\begin{array}{c}
		\mathbf{X} \\ \mathbf{X}^{\mathrm{kr}} \\ \mathbf{U} \\ \mathbf{U}^{\mathrm{kr}}
	\end{array}\right].
\end{equation}
Based on this, the full-order parametric matrices of the polynomial-dependent LPV model can be derived as
\begin{equation}
	\left[\mathcal{A} \quad \mathcal{B} \right] = \mathbf{X}^{+}\left[\begin{array}{c}
		\mathbf{X} \\ \mathbf{X}^{\mathrm{kr}} \\ \mathbf{U} \\ \mathbf{U}^{\mathrm{kr}}
	\end{array}\right]^{\dagger}
\end{equation}
where, $\dagger$ denotes pseudo-inverse. Therefore, the full-order model~\eqref{FOM_ss} and LPV full-order model can be denoted by a data-driven parametric representation in~\eqref{data_representation}.  

\begin{equation}\label{data_representation}
	\mathbf{x}_{k+1} =  \mathbf{X}^{+}\left[\begin{array}{c}
		\mathbf{X} \\ \mathbf{X}^{\mathrm{kr}} \\ \mathbf{U} \\ \mathbf{U}^{\mathrm{kr}}
	\end{array}\right]^{\dagger} \left[ \begin{array}{c}
		\mathbf{x}_k \\
		\Theta_k \otimes \mathbf{x}_k \\
		\mathbf{u}_k \\
		\Theta_k \otimes \mathbf{u}_k
	\end{array}\right]
\end{equation}

\subsection{Model Order Reduction by p-DMD}


Taking the singular value decomposition (SVD) on the data snapshots, one can obtain the model approximation by keeping the $n_{z}$ numbers of most significant singular values in $\Sigma_{r}$,
\begin{equation}
	\left[	\mathbf{X} ; \mathbf{X}^{\mathrm{kr}};  \mathbf{U};  \mathbf{U}^{\mathrm{kr}}\right] = \mathbf{U}\Sigma \mathbf{V}^{T} \approx \mathbf{U}_{r}\Sigma_{r} \mathbf{V}_{r}^{T}.
\end{equation}
Therefore, the full-order parametric matrices can be represented by 
\begin{equation}
	\left[\mathcal{A}, \mathcal{B}\right] = \mathbf{X}^{+} \mathbf{V}_{r}\Sigma_{r}^{-1}\mathbf{U}_{r}^{T}
\end{equation}
From the SVD of one-step shifted states sequence, the projection basis can be obtained by extracting the dominant singular values and the corresponding orthonormal vectors $\hat{U}_{z}$.
\begin{equation}
	\mathbf{X}^{+} = \hat{\mathbf{U}}\hat{\Sigma}\hat{\mathbf{V}}^{T} \approx \hat{\mathbf{U}}_{z}\hat{\Sigma}_{z}\hat{\mathbf{V}}_{z}^{T}
\end{equation}
Therefore, the reduced-order matrices are obtained from the projection basis from $\hat{U}_{z}$ is 
\begin{equation}\label{parametric_ROM_projection}
	\left[ \mathcal{A}_{r}, \mathcal{B}_{r}, \mathbf{C}_{r}\right] = \left[\hat{\mathbf{U}}_{z}^{T}\mathcal{A}\hat{\mathbf{U}}_{z}, \hat{\mathbf{U}}_{z}^{T}\mathcal{B}, \mathbf{C}\hat{\mathbf{U}}_{z} \right]. 
\end{equation}
where, $\mathcal{A}_{r} = \left[\mathbf{A}_{r0}, \mathbf{A}_{r1}, \cdots, \mathbf{A}_{rn_{p}}\right], \mathcal{B}_{r} = \left[\mathbf{B}_{r0}, \mathbf{B}_{r1}, \cdots, \mathbf{B}_{rn_{p}}\right]$. With the obtained matrices $\mathcal{A}_{r}, \mathcal{B}_{r}$, the reduced-order parameter-dependent system matrices can be extracted as 
\begin{equation}
	\mathbf{A}_{r}\left(\theta_k\right)=\mathbf{A}_{r0}+\sum_{i=1}^{n_{\mathrm{p}}} (\theta_k)^{i} \mathbf{A}_{ri}, \quad \mathbf{B}_{r}\left(\theta_k\right)=\mathbf{B}_{r0}+\sum_{i=1}^{n_{\mathrm{p}}} (\theta_k)^{i} \mathbf{B}_{ri}. 
\end{equation}

\begin{remark}
	The data snapshots $\left(\mathbf{U}, \mathbf{U}^{kr},\mathbf{X}, \mathbf{X}^{kr}, \mathbf{X}^{+} \right)$ are not required to be around the steady-state (equilibrium) flight condition and small deformation. Therefore, p-DMD has improved capability than traditional DMD~\cite{iannelli2021balanced,fonzi2020data} in terms of varying flight conditions (scheduling parameters) and large structure deformation. Besides, the transient dynamics caused by varying scheduling parameters can be captured by p-DMD as long as the data snapshot is obtained at dynamic flight conditions, which is studied and verified in Section \ref{sec_numericalstudy2}.
\end{remark}

\begin{remark}[state consistency]
	A constant projection vector $\hat{\mathbf{U}}_{z}$ is used for order reduction projection from full-order states, so state consistency is inherently guaranteed. The proposed method avoids performing DMD at multiple operating points and interpolating, so eigen-tracking in interpolation to obtain state consistency is not needed. On the contrary, the traditional DMD cannot always render system matrices with state consistency across multiple fixed scheduling parameters unless the same projection basis can be successfully found and implemented \cite{fonzi2020data,iannelli2021balanced}. Besides, the projection basis is formulated from snapshot of states, so the LPV-ROM's states are physically interpretative as the dominant behaviors of states. 
\end{remark}

\section{Numerical Study Case 1: Highly Flexible Cantilever Wing}\label{sec_numericalstudy1}

The highly flexible rectangular wing provided in \cite{patil2001nonlinear} is studied for the current study. The physical and geometrical properties of the wing are listed in Table \ref{tab: wingprop}. It is divided into ten elements with a clamped root. A trailing-edge control surface on the wing is defined from its \SI{60}{\percent} to \SI{90}{\percent} span, occupying \SI{20}{\percent} of the chord.

\begin{table}[htbp]
	\centering
	\caption{Properties of a highly flexible wing}
	\begin{tabular}{ll}
		\toprule
		\toprule
		Quantity & Value \\ 
		\midrule 
		Length & \qty{16}{\meter} \\ 
		Chord & \qty{1}{\meter} \\ 
		Spanwise ref.\ axis location (from L.E.) & \qty{50}{\percent} of chord \\ 
		Center of gravity (from L.E.) & \qty{50}{\percent} of chord \\ 
		Torsional rigidity & \qty{1e4}{\newton\meter\squared} \\ 
		Flat bending rigidity & \qty{2e4}{\newton\meter\squared} \\ 
		Chord bending rigidity & \qty{4e6}{\newton\meter\squared} \\ 
		Mass per unit length & \qty{0.75}{\kilogram\per\meter} \\ 
		Rotational Inertia per unit length & \qty{0.1}{\kilogram\meter} \\ 
		\bottomrule
		\bottomrule
	\end{tabular}%
	\label{tab: wingprop}%
\end{table}%

\subsection{Fixed Scheduling Parameter (Angle of Attack)}

The flexible wing is placed in a freestream of $U_{\infty} = \qty{20}{\meter\per\second}$ at the altitude of \qty{20000}{\meter}. The angle of attack measured at the wing root is selected as the scheduling parameter for LPV models. The models span at grid angles of attack ranging from $-10^{\circ}$ to $10^{\circ}$ with an increment of $0.5^{\circ}$, resulting in 41 grid models. The states of the full-order model include four strains, four strain rates, and six aerodynamic states for each beam element. Therefore, the full-order state-space model has a dimension of 140. The outputs are selected as the out-of-plane bending strains of all ten elements. Since it is cantilevered at the root, the rigid-body dynamics are not included in the study of the slender wing.

\begin{figure}[ht]
	\centering
	\includegraphics[trim={0.5in 3in 0.45in 3.2in},clip,width=0.5\linewidth]{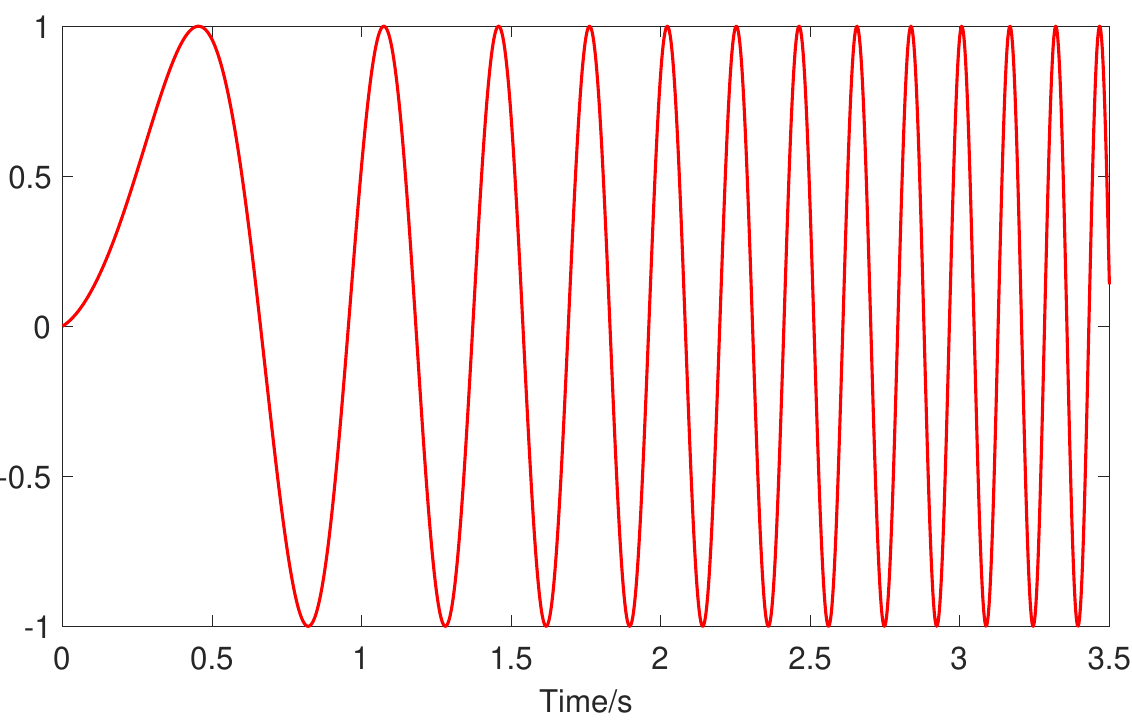}
	\caption{Excitation input of a chirp signal with sweep frequency from 0.1-10 Hz.}
	\label{fig:chirpinput}
\end{figure}

The setup of simulations is as follows to obtain the data snapshots at an equilibrium point of fixed angle of attack. A chirp signal in Fig.~\ref{fig:chirpinput} is applied to the main-wing flap at one fixed angle of attack to excite the aeroelastic modes, and the data snapshot of states and inputs are collected to conduct the data-driven LPV-ROM modeling algorithm. The chirp signal consists of frequency components spanning from $0.1$ Hz to $10$ Hz. The full-order linearized aeroelastic system is simulated at the sampling rate of $T_{s} = \qty{0.001}{\second}$. 

\subsubsection{Model order versus modeling accuracy}

\begin{figure}[h!]
	\centering
	\includegraphics[width=0.65\linewidth]{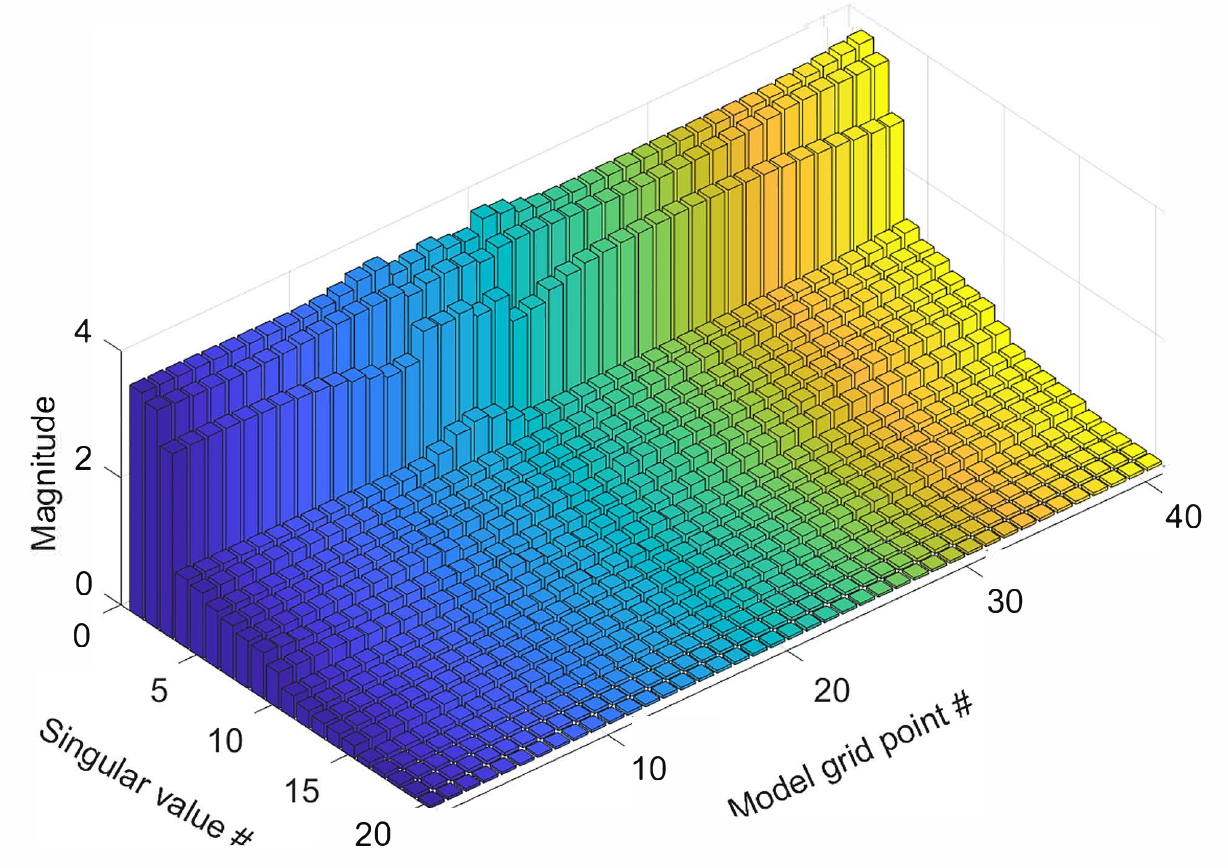}
	\caption{The 20 largest singular values of grid models at fixed scheduling parameters.}
	\label{fig:singularvaluesmodelindex}
\end{figure}

The p-DMD method is repeatedly applied at grid angles of attack, and the order of ROMs versus modeling accuracy is explored. The first 20 most significant singular values $\Sigma$ are plotted in Fig. \ref{fig:singularvaluesmodelindex}. It can be seen the first 12 singular values can retain over $95 \%$ of the summation of all singular values, and is selected as the order of ROMs.

\begin{figure}[h!]
	\centering
	\includegraphics[trim={0in 2.5in 0in 2.6in},clip,width=0.75\linewidth]{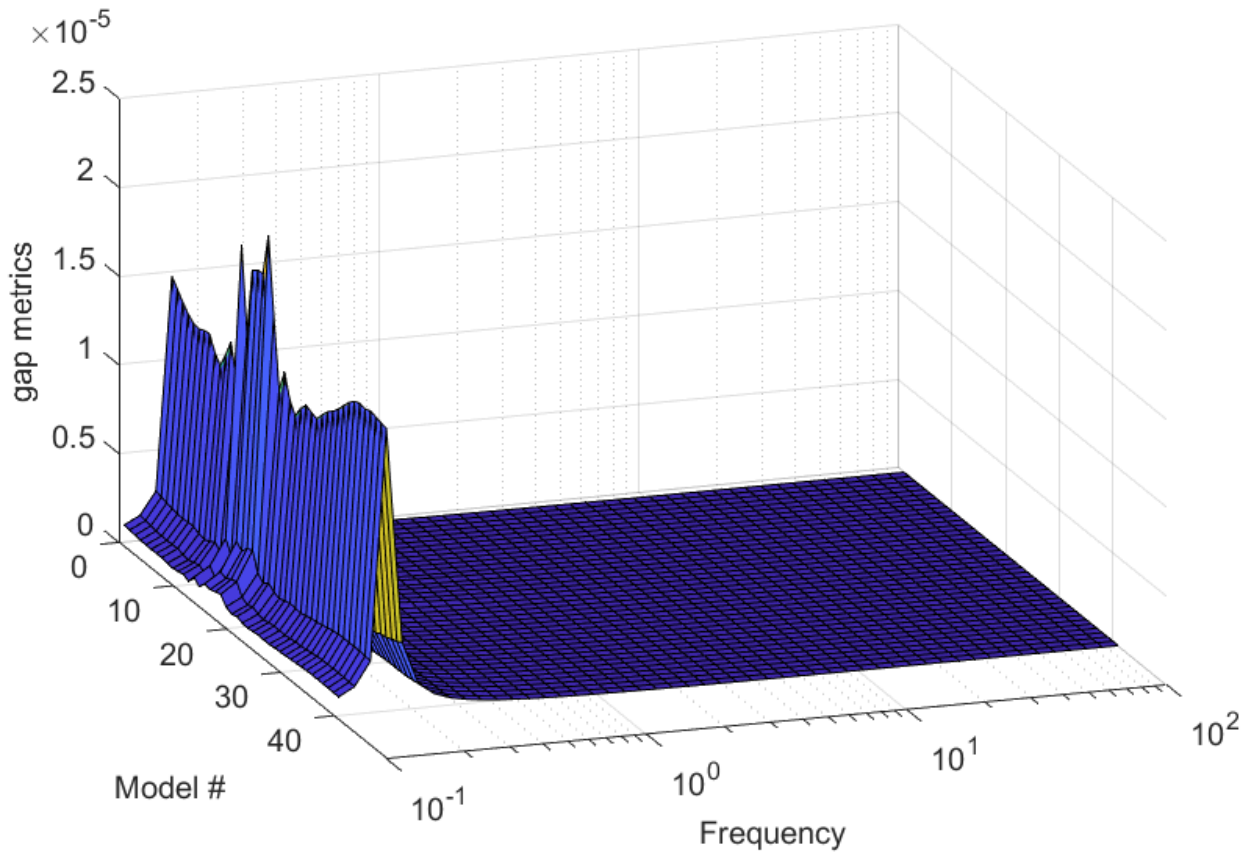}
	\caption{$v$-gap metric surface of frequency responses at fixed models. Gap metric close to 0 meaning that two systems are close.}
	\label{fig:vgaplpvromfom}
\end{figure}

The $v$-gap metrics between the ROMs and FOMs at fixed parameters are plotted in Fig.~\ref{fig:vgaplpvromfom} to evaluate the model accuracy of the ROMs. The gap metric indicates the distances between two systems and ranges between 0 and 1. A value close to 0 means that the FOM and ROM are close, and the stabilizing controller of ROM can stabilize FOM. Fig.~\ref{fig:vgaplpvromfom} shows the maximum gap metric occurs around $0.1$ Hz, and has a very small value of $10^{-5}$ level. Therefore, the LPV-ROM model of order $12$ can accurately capture the full-order model in the selected frequency range.

\begin{figure}[ht!]
	\centering
	\begin{subfigure}[b]{0.75\linewidth}
		\centering
		\includegraphics[trim={0in 4.25in 0in 4.25in},clip,width=\linewidth]{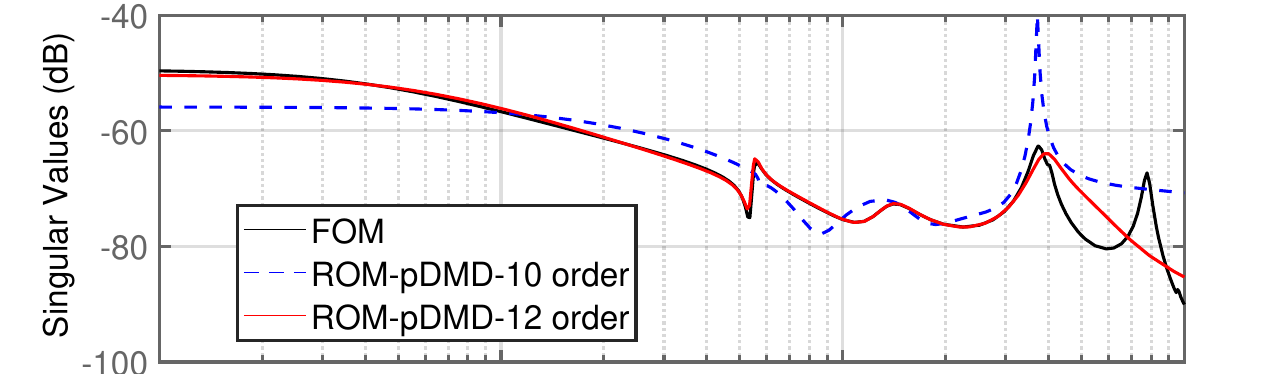}
		\label{fig:ModelIndex1_Order10_Output1_freq}
	\end{subfigure}
	\begin{subfigure}[b]{0.75\linewidth}
		\centering
		\vspace{-1em}
		\includegraphics[trim={0in 4.25in 0in 4.25in},clip,width=\linewidth,]{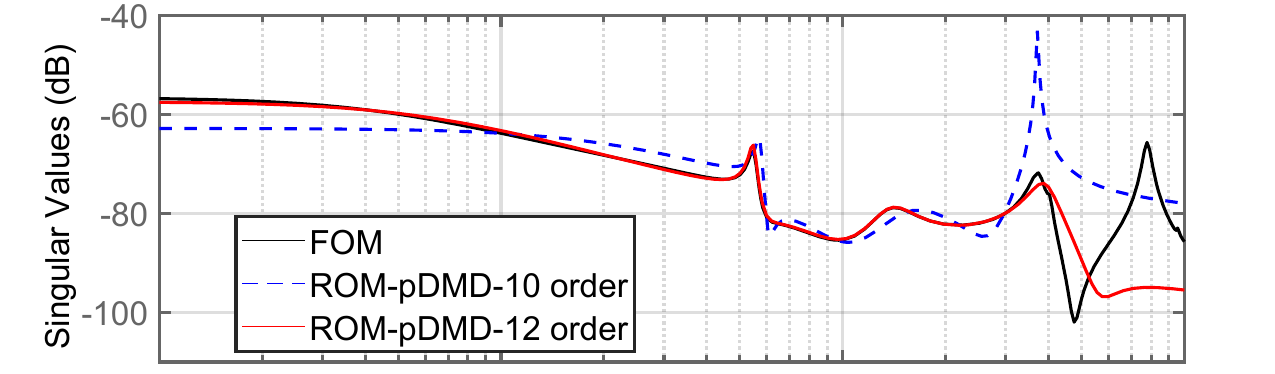}
		\label{fig:ModelIndex1_Order12_Output1_freq}
	\end{subfigure}

	\begin{subfigure}[b]{0.75\linewidth}
		\centering
		\vspace{-1em}
		\includegraphics[trim={0in 3.7in 0in 4.25in},clip,width=\linewidth,]{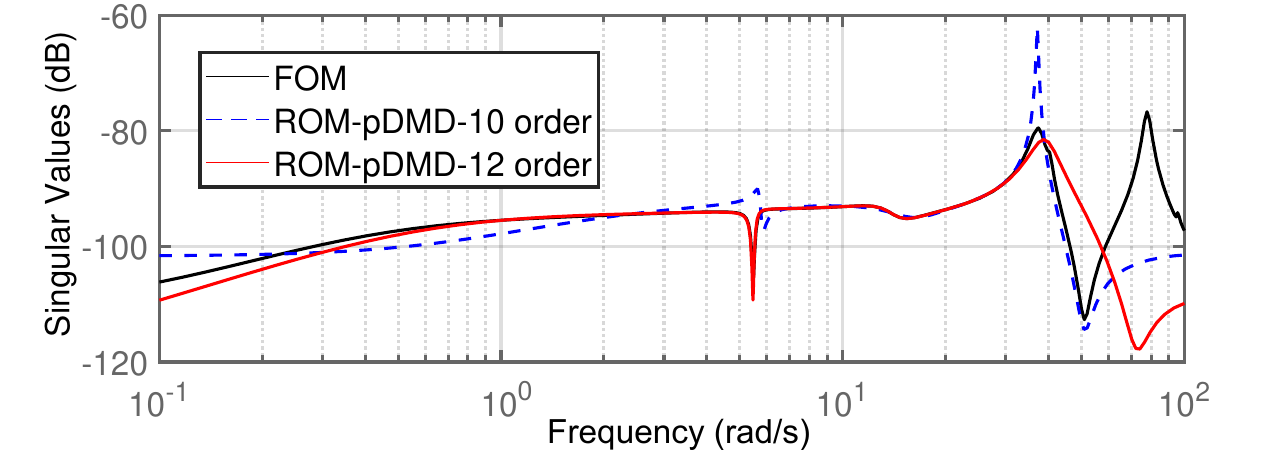}
		\label{fig:ModelIndex1_Order14_Output1_freq}
	\end{subfigure}
	\caption{Frequency responses of $u$ to $y_{1}$ (top), $y_{5}$ (middle) and $y_{10}$ (bottom), with 10$^{\mathrm{th}}$, 12$^{\mathrm{th}}$-order ROMs at angle of attack $\alpha = -10^{\circ}$.}
	\label{fig:Modelidx1_orders_freq}
\end{figure}

\begin{figure}[ht!]
	\centering
	\begin{subfigure}[b]{0.75\linewidth}
		\centering
		\vspace{-1em}
		\includegraphics[trim={0in 4.55in 0in 4.0in},clip,width=\linewidth]{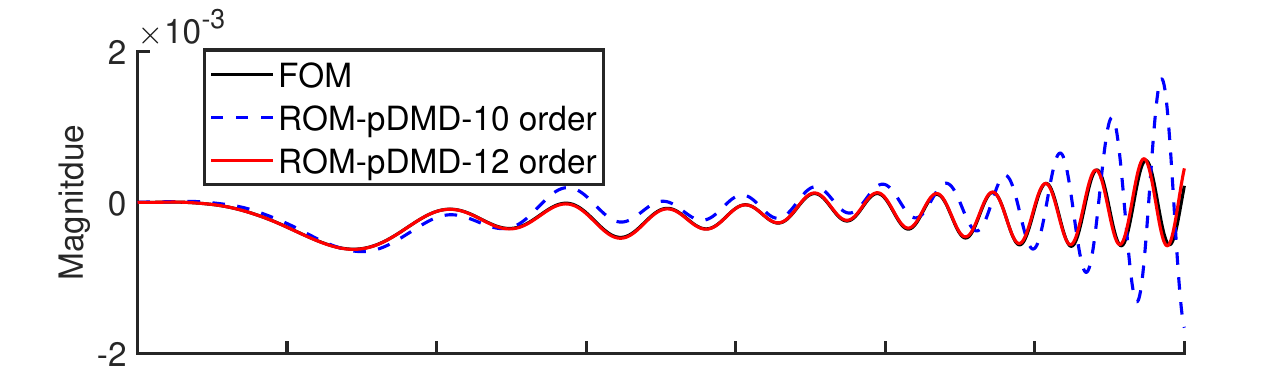}
		\label{fig:ModelIndex1_Order10_Output1_time}
	\end{subfigure}
	
	\begin{subfigure}[b]{0.75\linewidth}
		\centering
	    \vspace{-1em}
		\includegraphics[trim={0in 4.55in 0in 4.0in},clip,width=\linewidth]{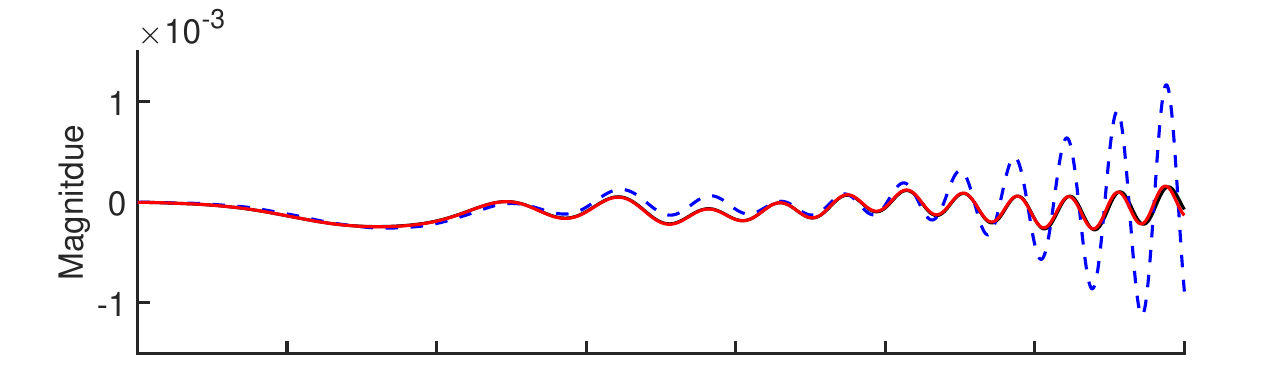}
		\label{fig:ModelIndex1_Order12_Output1_time}
	\end{subfigure}
	
	\begin{subfigure}[b]{0.75\linewidth}
		\centering
			    \vspace{-1em}
		\includegraphics[trim={0in 4.0in 0in 4.0in},clip,width=\linewidth]{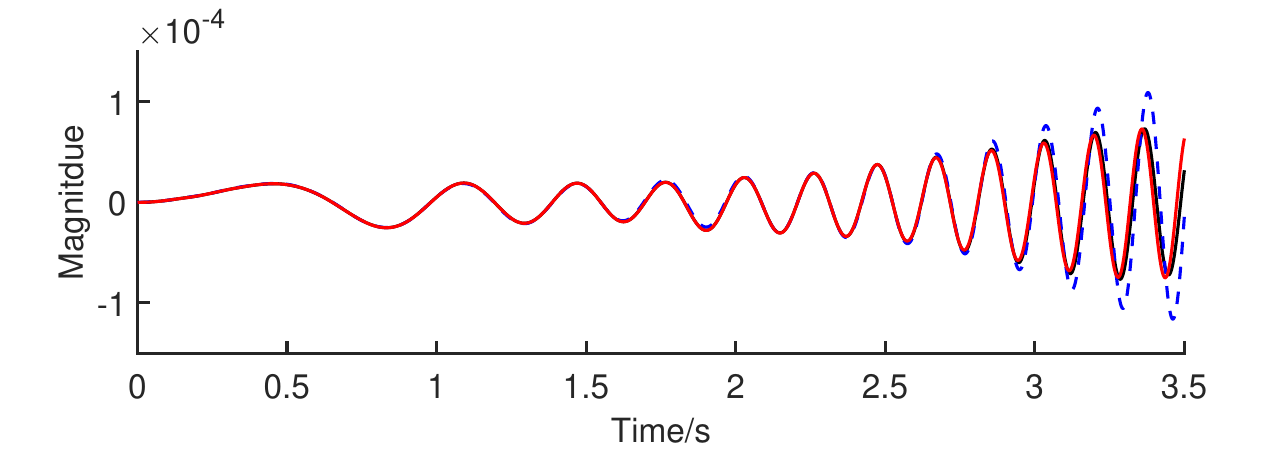}
		\label{fig:ModelIndex1_Order14_Output1_time}
	\end{subfigure}
	
	\caption{Time-domain responses of $y_{1}$ (top), $y_{5}$ (middle) and $y_{10}$ (bottom), with 10$^{\mathrm{th}}$, 12$^{\mathrm{th}}$-order ROMs at angle of attack $\alpha = -10^{\circ}$.}
	\label{fig:Modelidx1_orders_time}
\end{figure}

\begin{figure}[ht]
	\centering
	\begin{subfigure}[b]{0.75\linewidth}
		\centering
		\includegraphics[trim={0in 4.25in 0in 4.25in},clip,width=\linewidth,]{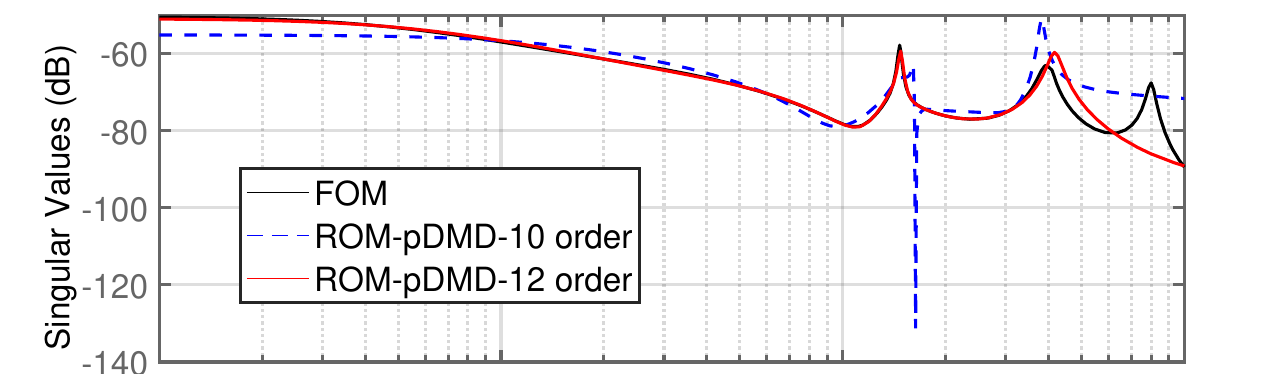}
		\label{fig:ModelIndex21_Order10_Output1_freq}
	\end{subfigure}
	
	\begin{subfigure}[b]{0.75\linewidth}
		\centering
		 \vspace{-1em}
		\includegraphics[trim={0in 4.25in 0in 4.25in},clip,width=\linewidth,]{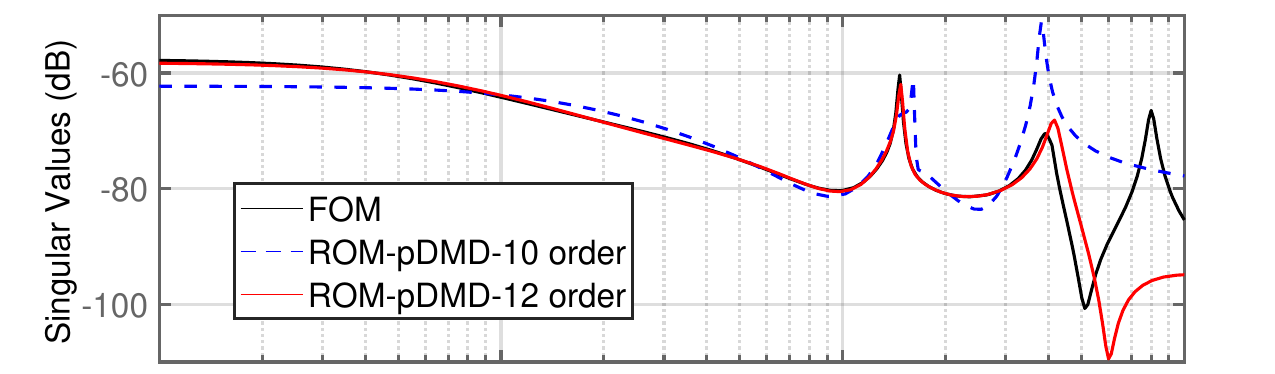}
		\label{fig:ModelIndex21_Order12_Output1_freq}
	\end{subfigure}
	
	\begin{subfigure}[b]{0.75\linewidth}
		\centering
		\vspace{-1em}
		\includegraphics[trim={0in 3.7in 0in 4.25in},clip,width=\linewidth,]{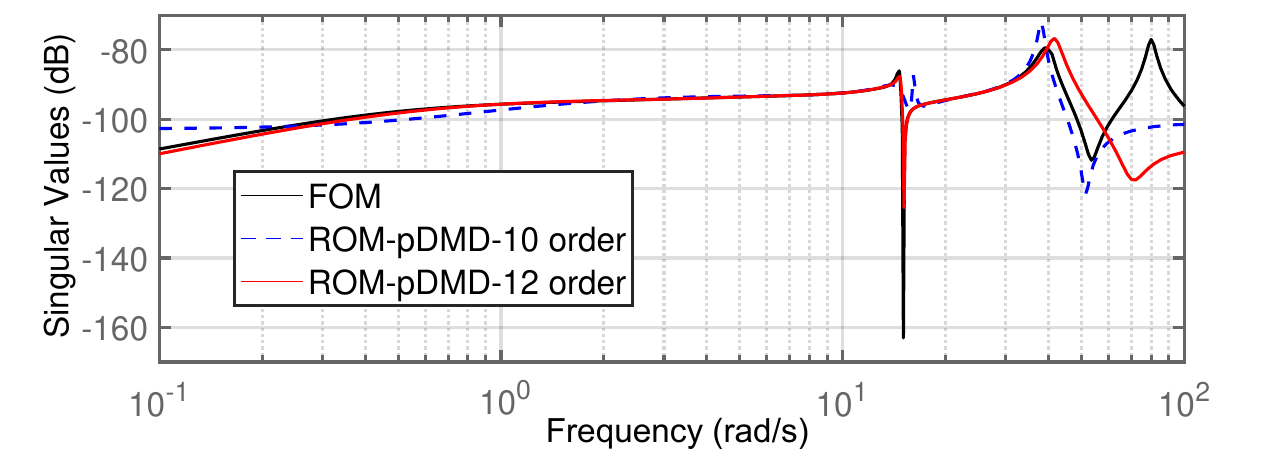}
		\label{fig:ModelIndex21_Order14_Output1_freq}
	\end{subfigure}
	\caption{Frequency responses of $u$ to $y_{1}$ (top), $y_{5}$ (middle) and $y_{10}$ (bottom), with 10$^{\mathrm{th}}$, 12$^{\mathrm{th}}$-order ROMs at angle of attack $\alpha = 0^{\circ}$.}
	\label{fig:Modelidx21_orders_freq}
\end{figure}

\begin{figure}[ht]
	\centering

	\begin{subfigure}[b]{0.75\linewidth}
		\centering
		\vspace{-1em}
		\includegraphics[trim={0in 4.55in 0in 4.0in},clip,width=\linewidth]{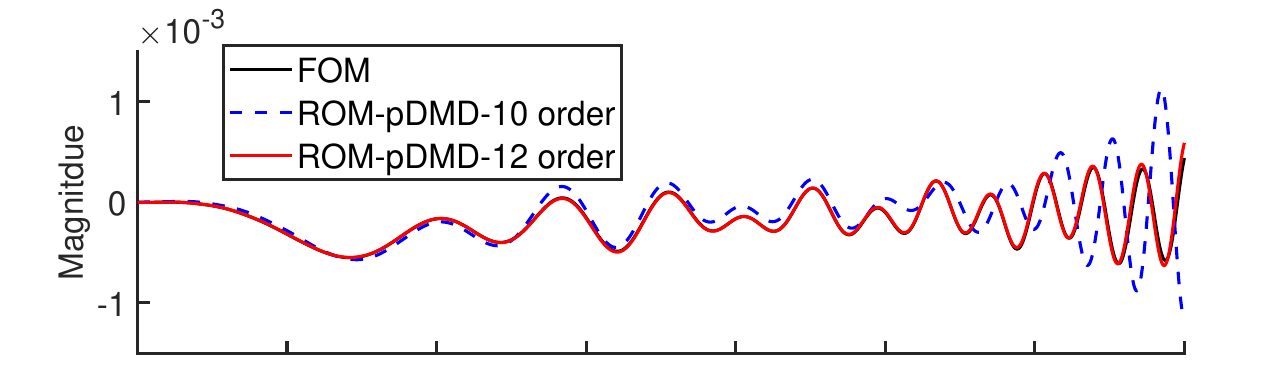}
		\label{fig:ModelIndex21_Order10_Output1_time}
	\end{subfigure}
	
	\begin{subfigure}[b]{0.75\linewidth}
		\centering
		\vspace{-1em}
		\includegraphics[trim={0in 4.55in 0in 4.0in},clip,width=\linewidth]{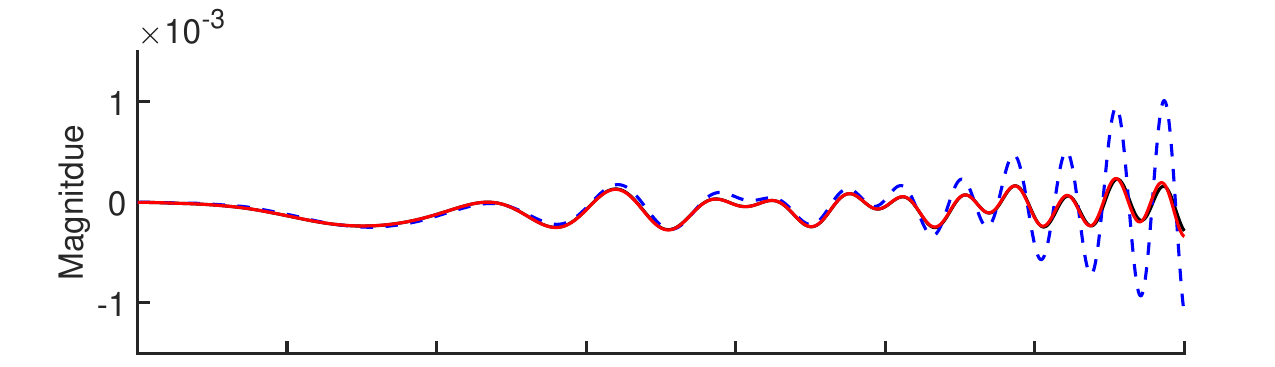}
		\label{fig:ModelIndex21_Order12_Output1_time}
	\end{subfigure}
	
	\begin{subfigure}[b]{0.75\linewidth}
		\centering
		\vspace{-1em}
		\includegraphics[trim={0in 4.55in 0in 4.0in},clip,width=\linewidth]{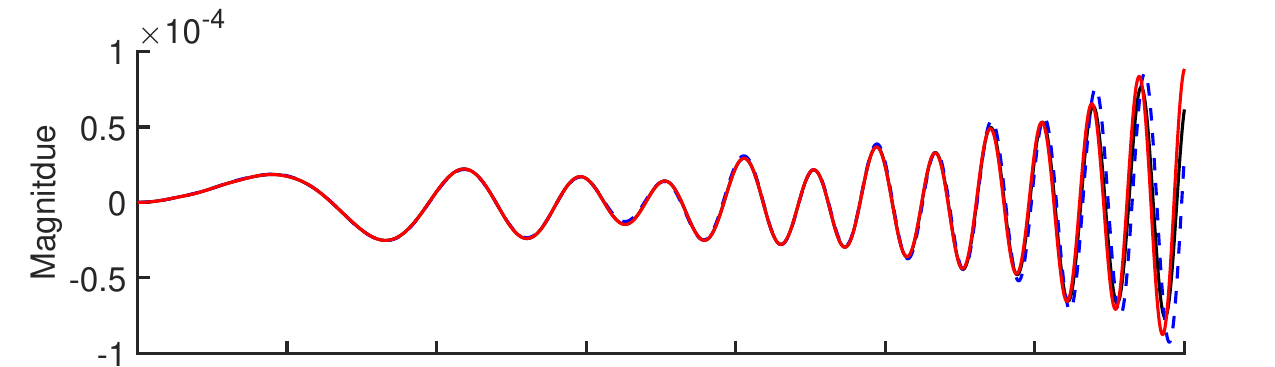}
		\label{fig:ModelIndex21_Order14_Output1_time}
	\end{subfigure}
	
\caption{Time-domain responses of $y_{1}$ (top), $y_{5}$ (middle) and $y_{10}$ (bottom), with 10$^{\mathrm{th}}$, 12$^{\mathrm{th}}$-order ROMs at angle of attack $\alpha = 0^{\circ}$.}
	\label{fig:Modelidx21_orders_time}
\end{figure}

At the angles of attack $\alpha$ equals \ang{-10} and \ang{0}, respectively, the time- and frequency-domain responses of linearized FOM, 10-order LPV-ROM, and 12-order LPV-ROM are compared in Figs.~\ref{fig:Modelidx1_orders_freq}-\ref{fig:Modelidx21_orders_time}. Note that, the plotted variables are the perturbed magnitudes from the trim points. It can be easily observed that the 10-order LPV-ROM has an obvious modeling error in the frequency range of concern, while the 12-order LPV-ROM model can capture the FOM in the frequency range of \qtyrange{0}{40}{rad/s} ($\approx 6.3$ \unit{\hertz}). Therefore, the 12-order LPV-ROM is the ``minimum order'' model that can accurately describe the system dynamics. The polynomial order is selected as 4 to formulate LPV-ROM.

\subsubsection{Verification between p-DMD modeling and linearization-based modeling}

\begin{figure}[h!]
	\centering
	\vspace{-0em}
	\includegraphics[trim={1cm 7cm 1cm 6.5cm},clip,width=0.75\linewidth]{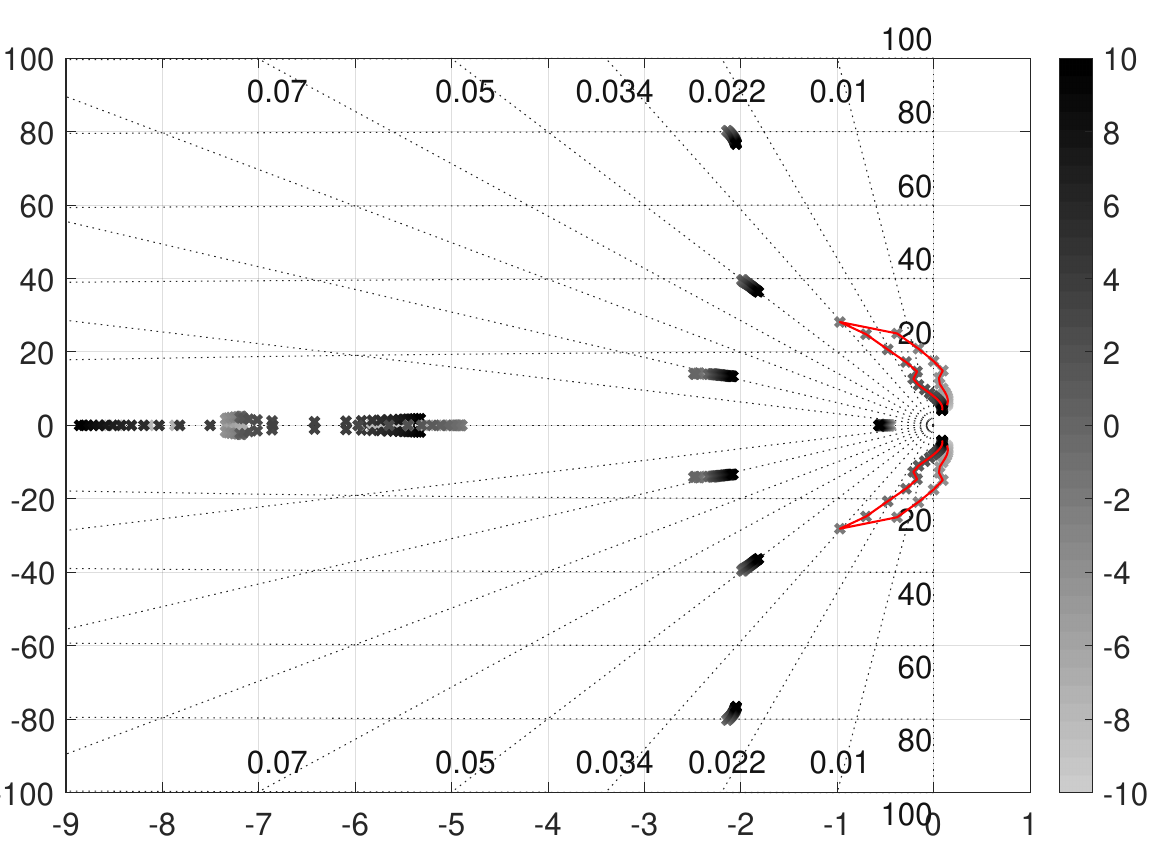}
	\caption{Root loci of ROMs of slender wing at grid points of angle of attack.}
	\label{fig: rootlocirom}
\end{figure}

In this section, the results from the authors' previous work of linearization-based modeling are compared with the data-driven p-DMD modeling. The nonlinear model~\eqref{eq: nl_ae_fd_eom} is linearized and reduced order at grid angles of attack to obtain a series of LTI-FOM. The root loci of the ROM are shown in Fig.~\ref{fig: rootlocirom}.

\begin{figure}[ht!]
	\centering
	\begin{subfigure}[b]{0.75\linewidth}
		\centering
		\includegraphics[width=\textwidth]{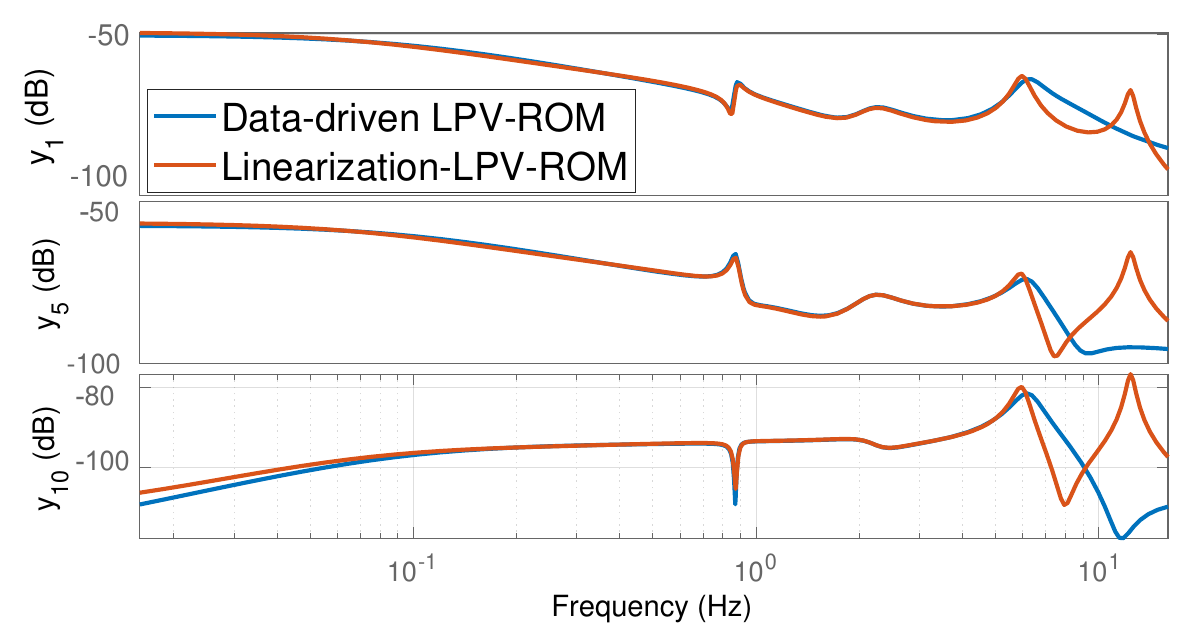}
	\end{subfigure}
	\hfill
	\begin{subfigure}[b]{0.75\linewidth}
		\centering
		\includegraphics[width=\textwidth]{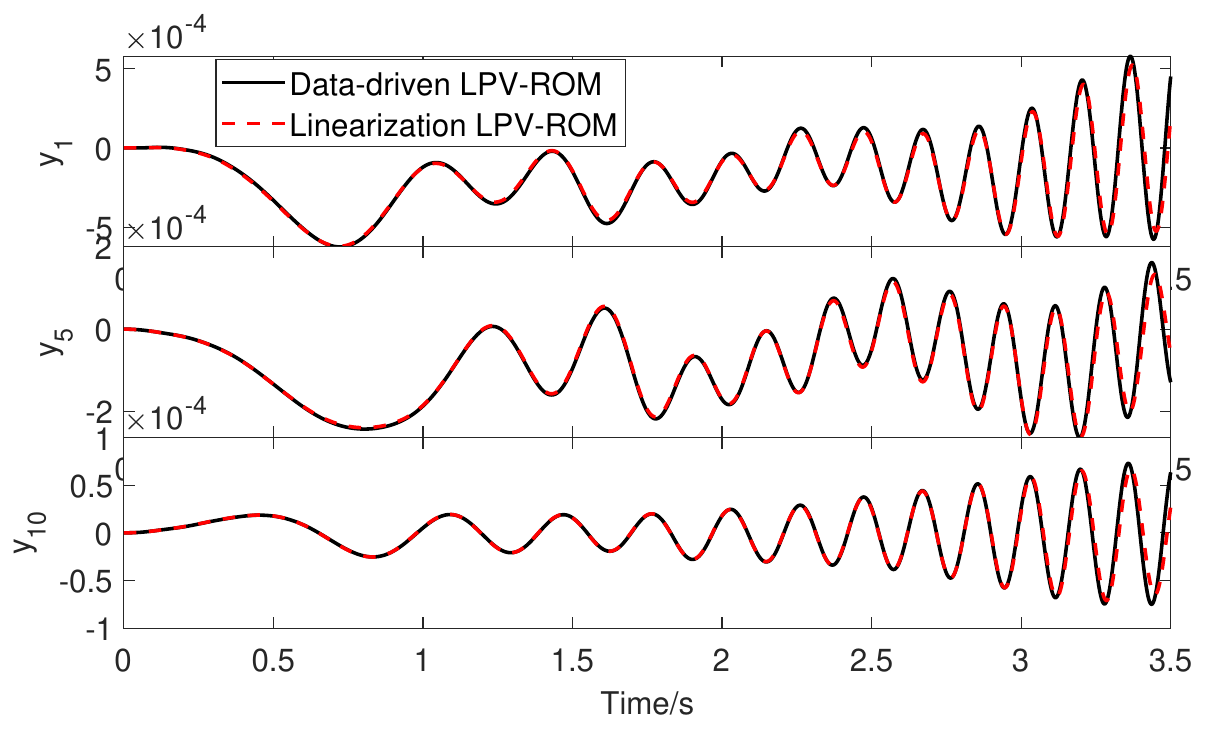}
	\end{subfigure}
\caption{Frequency and time-domain responses comparison of data-driven ROMs and linearization-based ROMs at $\alpha = -10^{\circ}$.}
	\label{fig:LM_data_comparison_Modelindex1}
\end{figure}

\begin{figure}[ht!]
	\centering
	\begin{subfigure}[b]{0.75\linewidth}
		\centering
		\includegraphics[width=\textwidth]{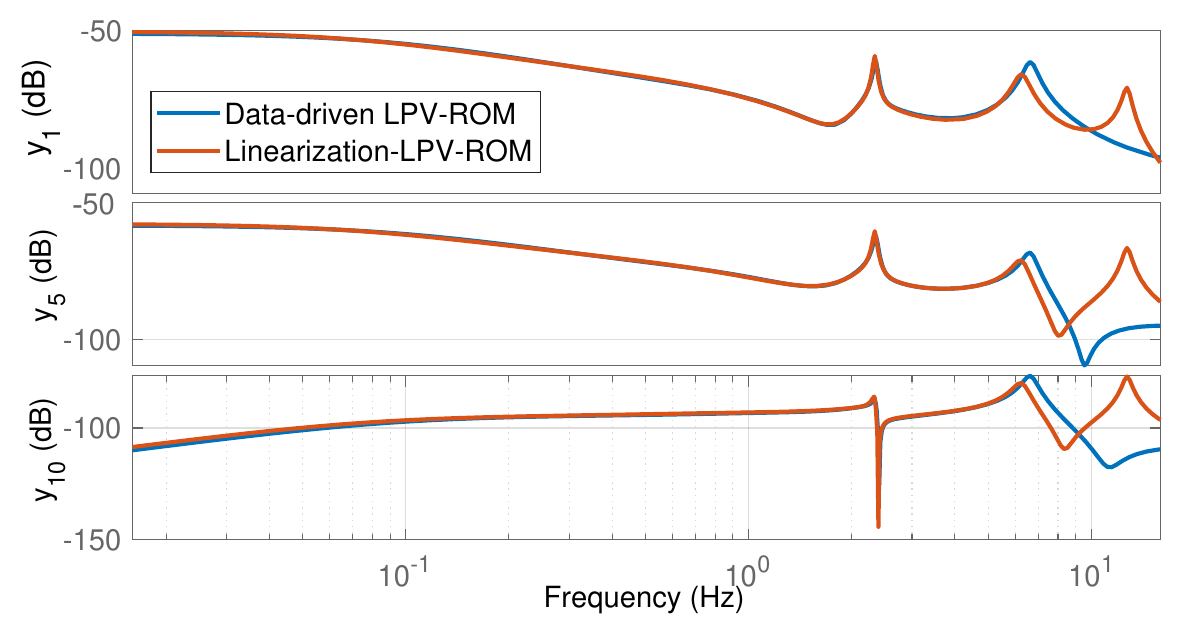}
	\end{subfigure}
	\hfill
	\begin{subfigure}[b]{0.75\linewidth}
		\centering
		\includegraphics[width=\textwidth]{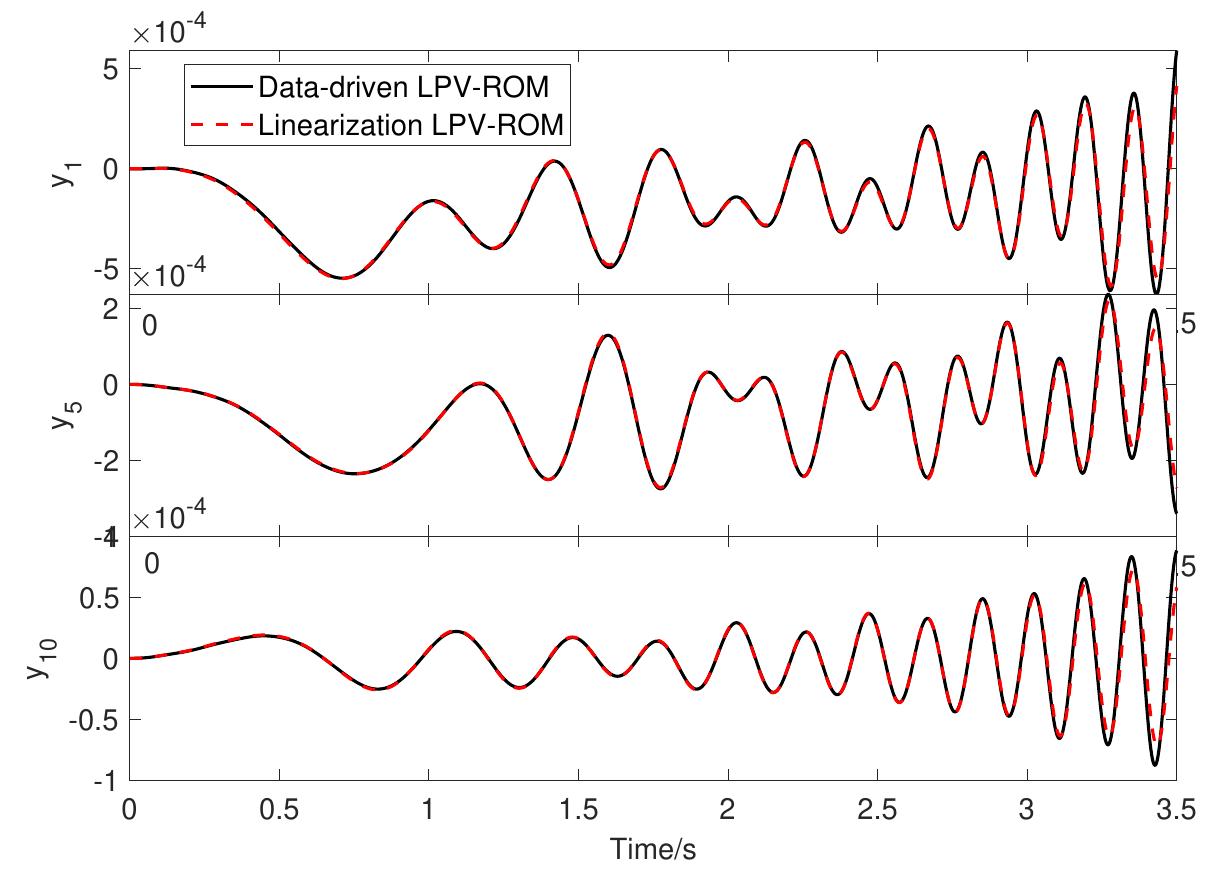}
	\end{subfigure}
	\caption{Frequency and time-domain responses comparison of data-driven ROMs and linearization-based ROMs at $\alpha = 0^{\circ}$.}
	\label{fig:LM_data_comparison_Modelindex21}
\end{figure}

For reasonable comparisons, the equilibrium conditions at multiple angles of attacks are focused. The data snapshot are obtained by simulating the aeroelastic model at small perturbations of steady state. The comparisons are plotted in Figs.~\ref{fig:LM_data_comparison_Modelindex1} and \ref{fig:LM_data_comparison_Modelindex21}. The data-driven LPV-ROM is found to match the linearization-based LPV-ROM, verifying that the data-driven approach can capture the physical characteristics of the aeroelasticity.

It is worth mentioning the scenarios of $\alpha = -10^{\circ}, + 10^{\circ}$, the eigenvalues have a positive real part (see root loci in Fig. \ref{fig: rootlocirom}), indicating that the aeroelastic system is unstable at these angles of attack. In the linearized-based approach, the $\sigma$-shifted $H_{2}$ norm is calculated by shifting the eigenvalue by $\sigma$ to obtain a stable system~\cite{al2017lpv}. However, the proposed p-DMD directly finds the dominant modes by analyzing the singular values of parametric Hankel matrices. The p-DMD approach can address the unstable dynamics. 

To conclude this section, the p-DMD method can accurately capture the local dynamics of the highly flexible wing at a fixed angle of attack, and the model accuracy is verified with the well-established linearization-based approach in frequency- and time-domain simulations.

\section{Numerical Study Case 2: Highly Flexible Aircraft}\label{sec_numericalstudy2}

\begin{figure}[ht!]
	\centering
	\includegraphics[width = 0.5\linewidth]{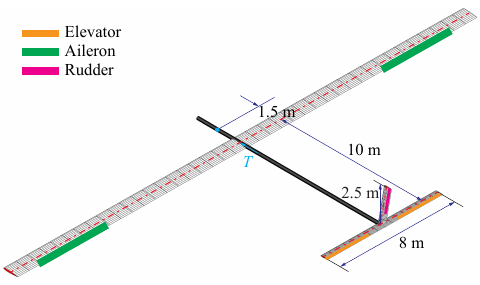}
	\caption{Platform of flexible vehicle.}
	\label{fig: flexvehicle_ailn}
\end{figure}

In this section, we study and show the results of data-driven modeling for a highly flexible aircraft with coupled elastic deformation and rigid-body motions. The aircraft is built based on the slender wing in Section \ref{sec_numericalstudy1}, as shown shown in Fig.~\ref{fig: flexvehicle_ailn}. All members of the aircraft are modeled as slender beams. However, the body and tail members are 100 times stiffer than the main wing, although they all have the same inertial properties. The chord length of the horizontal and vertical tails is \qty{0.5}{\meter}, while their beam reference lines are defined at a quarter of the chord from the leading edge. There is no aerodynamics coupled on the body. Similar to the main wing, trailing-edge control surfaces are also defined on the horizontal tail, from its \qty{25}{\percent} to the full span. The flaps on the horizontal tail are used as elevators to trim the vehicle for level flight and excite the longitudinal motion.

\subsection{Fixed Scheduling Parameter (Angle of Attack)}
In this simulation, the highly flexible aircraft is set cruising in a freestream of $U_{\infty} = \qty{22}{\meter\per\second}$ at the altitude of \qty{20000}{\meter}. For LPV modeling, the body angle of attack at the origin of the body frame is used as the scheduling parameter. The LPV models span at grid angles of attack ranging from $-7.5^{\circ}$ to $7.5^{\circ}$ with an increment of $0.5^{\circ}$. The total number of grid models is 31. The flexible vehicle model has 34 beam elements and four control inputs (two control surfaces of the main wing, one elevator, and one thrust). The symmetric wing vibrations and longitudinal rigid-body motion are considered. Instead of being used as ailerons, the surfaces on the main wing deflect symmetrically when excited. The FOM includes four strains, four strain rates, and six aerodynamic states for each beam element, in addition to six rigid-body velocities, four quaternions, and three inertial positions of the vehicle. Therefore, at a fixed angle of attack, the full-order $A$ matrix is of dimension $471 \times 471$, the $B$ matrix is of dimension $471 \times 3$, and the 26 outputs are selected as the out-of-plane bending strains on all main-wing elements (20 in total) plus longitudinal and vertical body velocities, pitch angle and rate, longitudinal and vertical body positions.

At a fixed angle of attack, the chirp signal in Fig.~\ref{fig:chirpinput} is fed as input perturbation of the main-wing flaps to excite the aeroelastic modes, and the data snapshots of states and inputs are collected to conduct the data-driven LPV-ROM modeling algorithm. The chirp signal consists of frequency components ranging from \qtyrange{0.1}{10}{\hertz}. The full-order linearized aeroelastic system is also simulated at the sampling rate of $T_{s} = \qty{0.001}{\second}$. 

\begin{figure}[h!]
	\centering
	\includegraphics[trim={0in 2.0in 0in 4.2in},clip,width=0.75\linewidth]{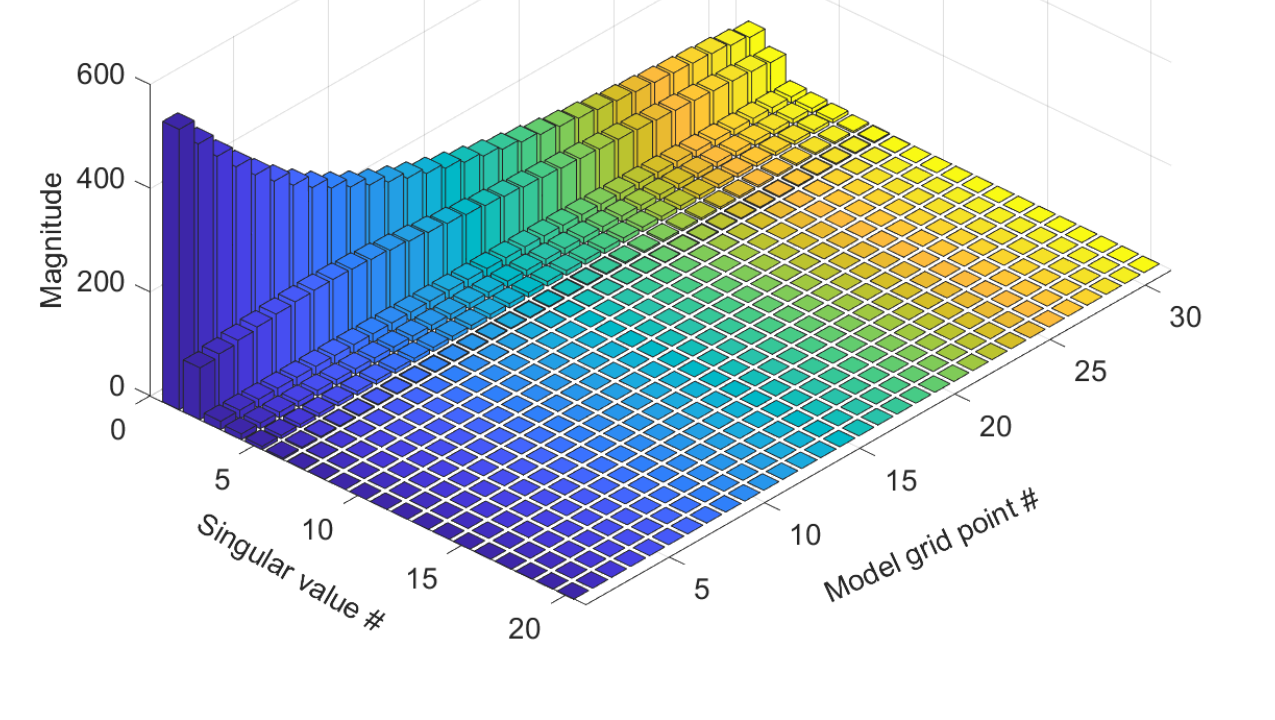}
	\caption{The 20 largest singular values of flexible aircraft at fixed scheduling parameters.}
	\label{fig:singularvalues_FlexVh}
\end{figure}

\begin{figure}[h!]
	\centering
	\includegraphics[trim={0cm 5cm 0cm 5.5cm},clip,width=0.75\linewidth]{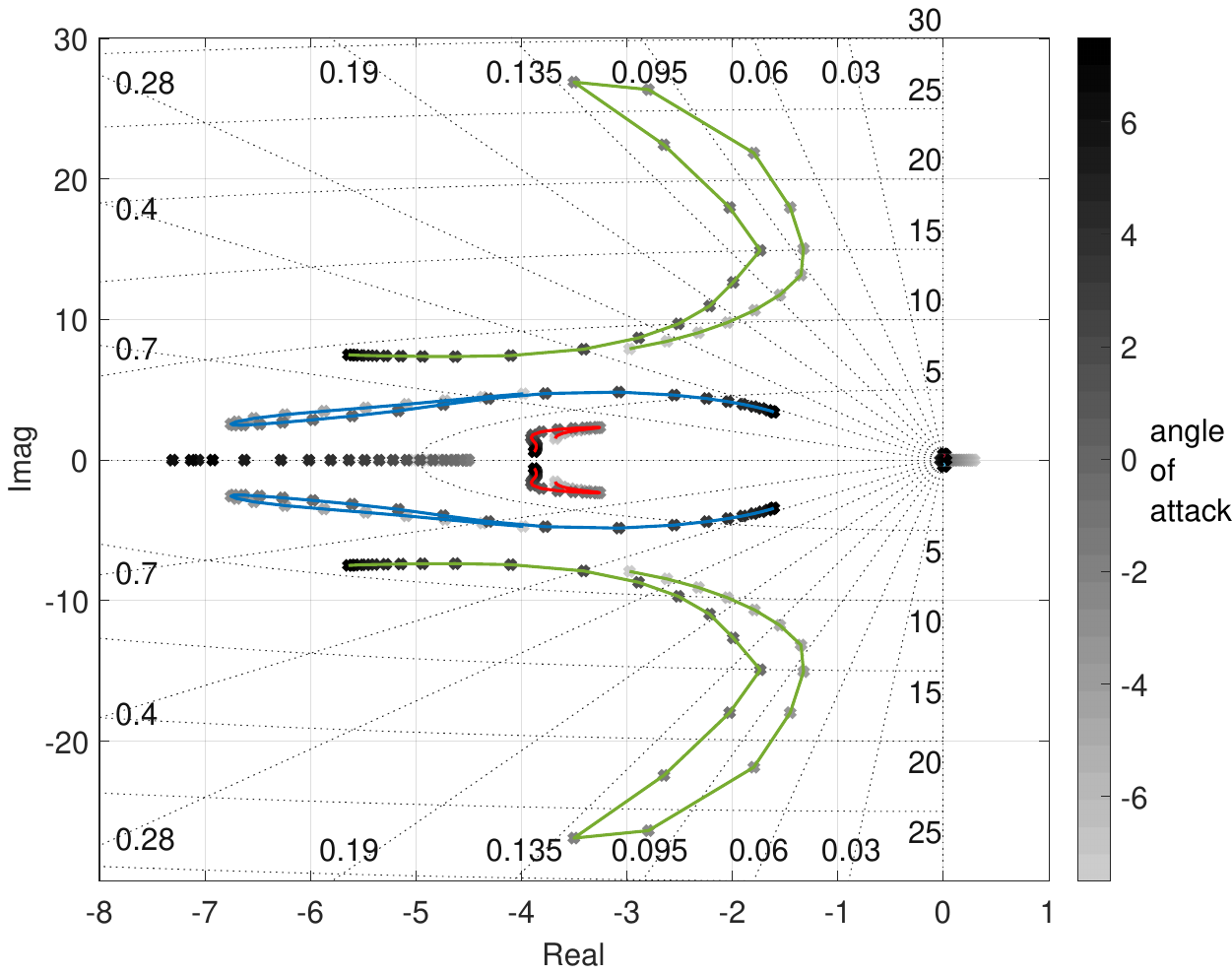}
	\caption{Root loci of ROMs of flexible aircraft at grid angles of attack.}
	\label{fig: FlexVh_root_loci_ROM}
\end{figure}

\begin{figure}[h!]
	\centering
	\includegraphics[trim={0in 2.0in 0in 3.5in},clip,width=0.75\linewidth]{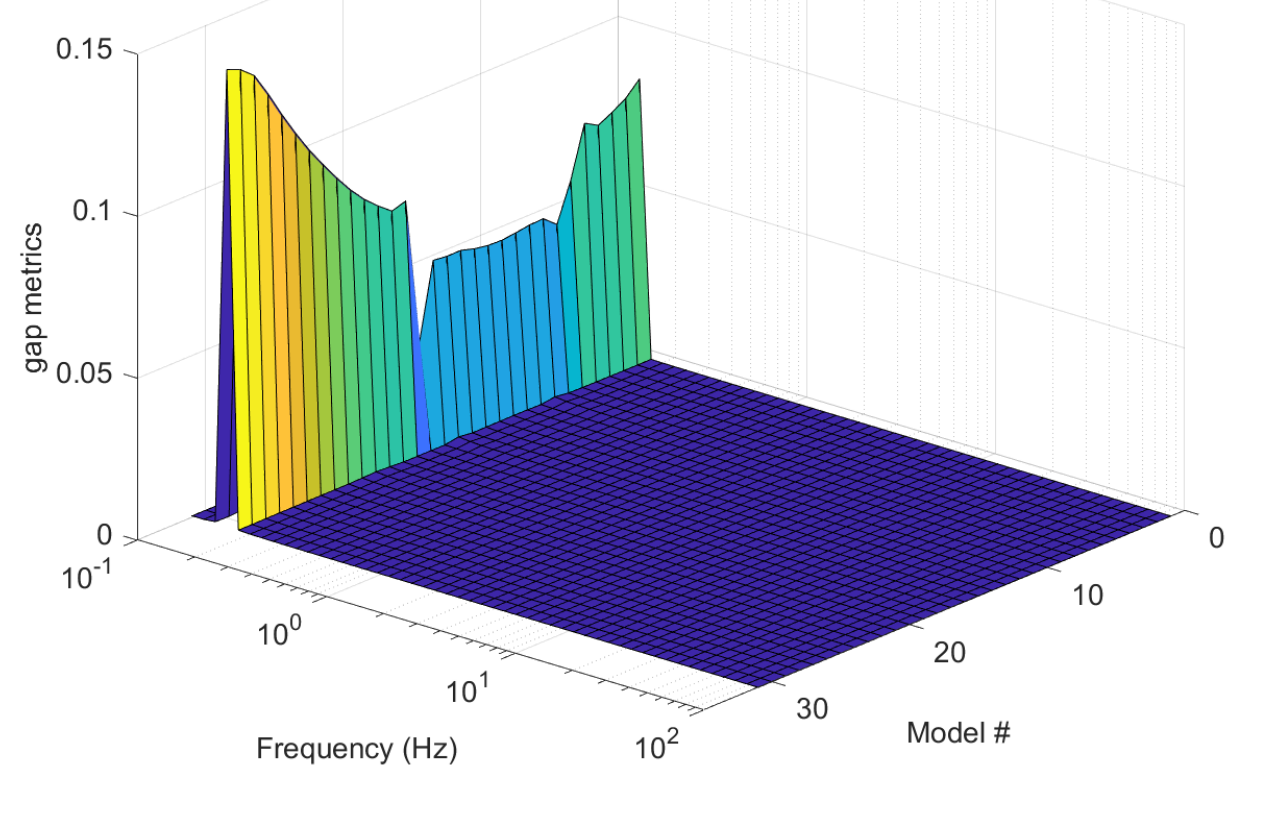}
	\caption{$v$-gap metric surface of frequency responses at fixed models. Gap metric close to 0 meaning that two systems are similar.}
	\label{fig:vgaplpvromfom_FlexVh}
\end{figure}

The 20 largest singular values of the state-Hankel matrix are plotted in Fig.~\ref{fig:singularvalues_FlexVh}, among which the 12 largest singular values are determined to be able to capture over $95\%$ accuracy of FOMs. The root loci of flexible vehicles at varying angles of attack is Fig.~\ref{fig: FlexVh_root_loci_ROM}, which is obtained from model reduction from FOMs. The v-gap metric comparison between the LTI-ROMs and LPV-ROMs by applying p-DMD on flexible vehicles is plotted in Fig.~\ref{fig:vgaplpvromfom_FlexVh}. These gap metrics are close to 0 at most of the regime, indicating the data-driven LPV-ROM can capture the local dynamics with excellent accuracy. 

\begin{figure}[ht]
	\centering
	\begin{subfigure}[b]{0.75\linewidth}
		\centering
		\includegraphics[trim={0.0in 4.25in 0in 4.0in},clip,width=\linewidth]{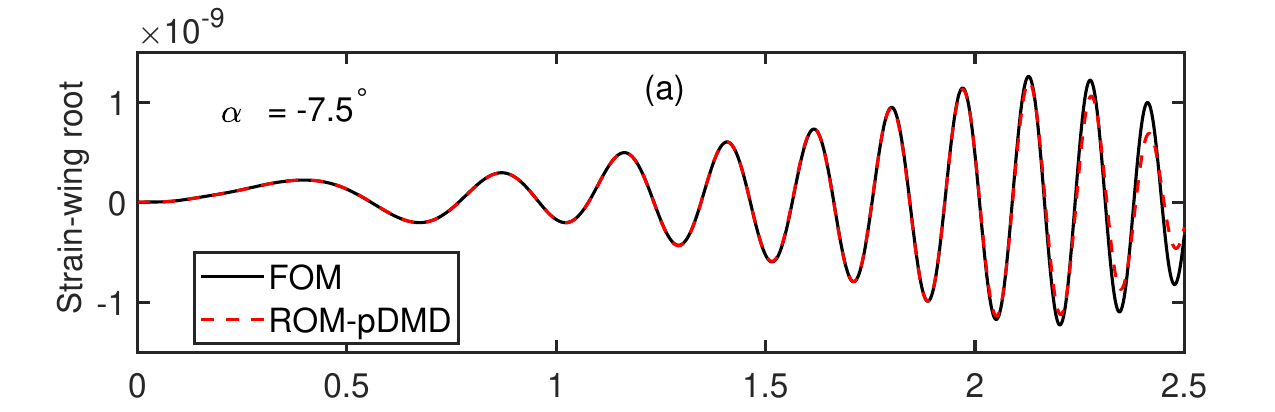}
		\label{fig:FlexVh_bendingstrain_wingroot_MoInd1}
	\end{subfigure}
	\hfill
	\begin{subfigure}[b]{0.75\linewidth}
		\centering
		\vspace{-3em}
		\includegraphics[trim={0.0in 4.25in 0in 4.0in},clip,width=\textwidth]{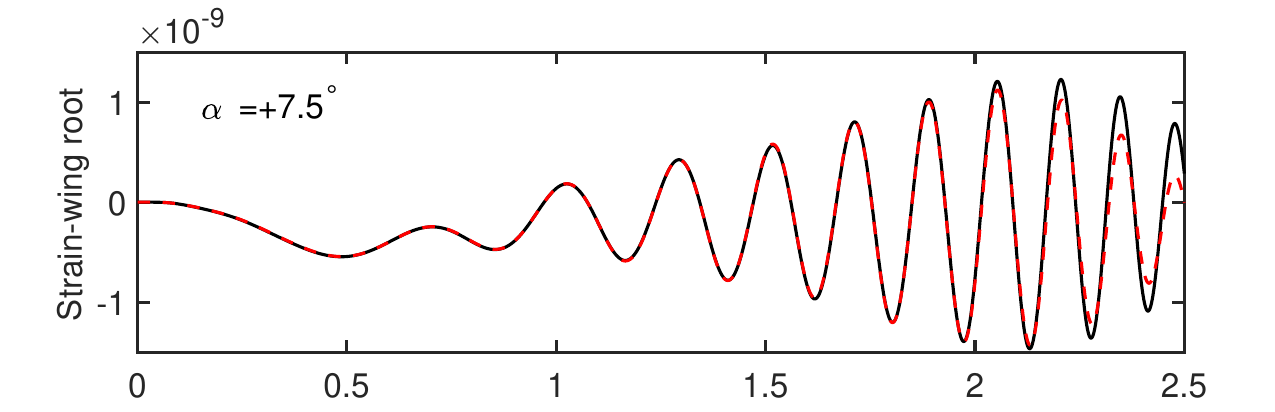}
		\label{fig:FlexVh_bendingstrain_wingroot_MoInd31}
	\end{subfigure}
	
	\begin{subfigure}[b]{0.75\linewidth}
		\centering
		\vspace{-3em}
		\includegraphics[trim={0.0in 4.35in 0in 4.0in},clip,width=\textwidth]{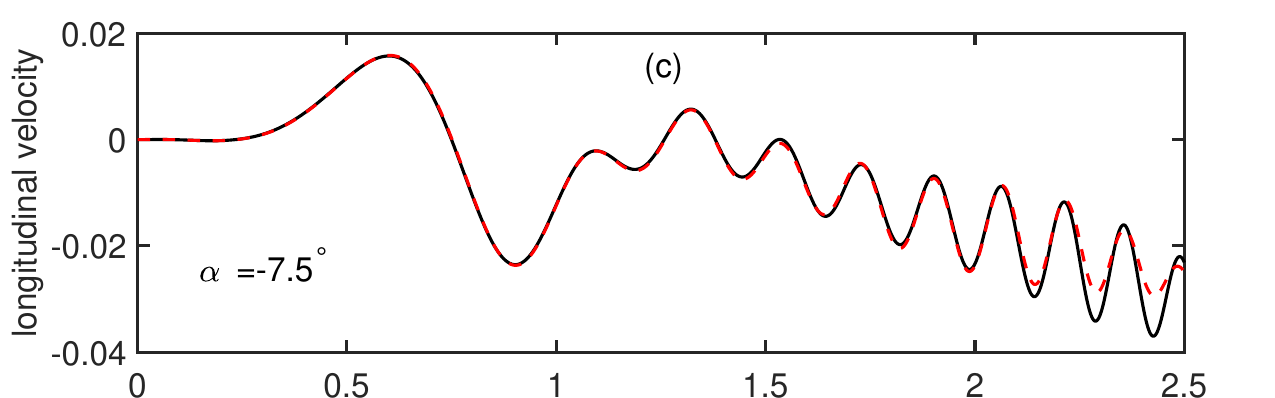}
		\label{fig:FlexVh_longitudinal_vel_MoInd1}
	\end{subfigure}
	\hfill
	\begin{subfigure}[b]{0.75\linewidth}
		\centering
			\vspace{-3em}
		\includegraphics[trim={0.0in 4.25in 0in 3.9in},clip,width=\textwidth]{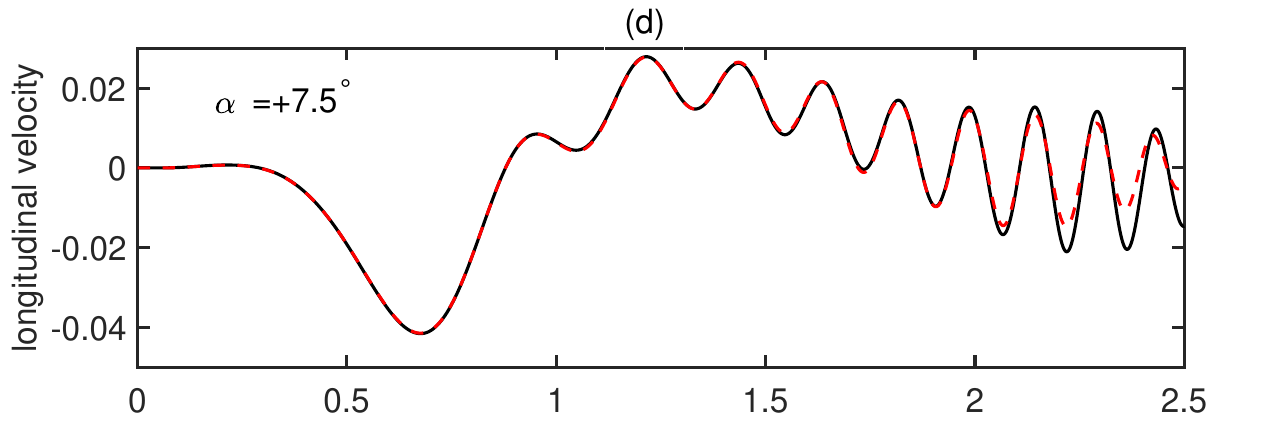}
		\label{fig:FlexVh_longitudinal_vel_MoInd31}
	\end{subfigure}
	
	\begin{subfigure}[b]{0.75\linewidth}
		\centering
		\vspace{-3em}
		\includegraphics[trim={0.0in 4.25in 0in 4.0in},clip,width=\textwidth]{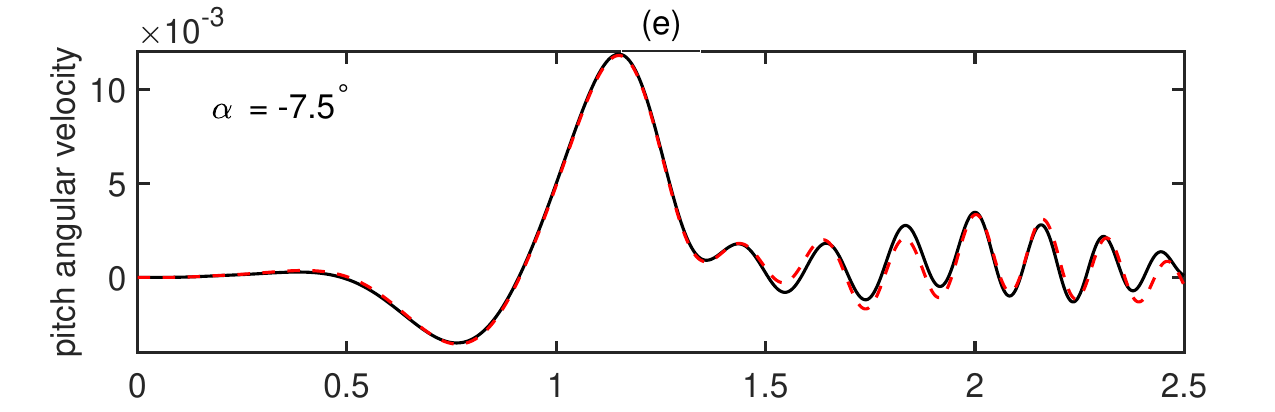}
		\label{fig:FlexVh_pitchrate_MoInd1}
	\end{subfigure}
	\hfill
	\begin{subfigure}[b]{0.75\linewidth}
		\centering
		\vspace{-3em}
		\includegraphics[trim={0.0in 4.25in 0in 4.0in},clip,width=\textwidth]{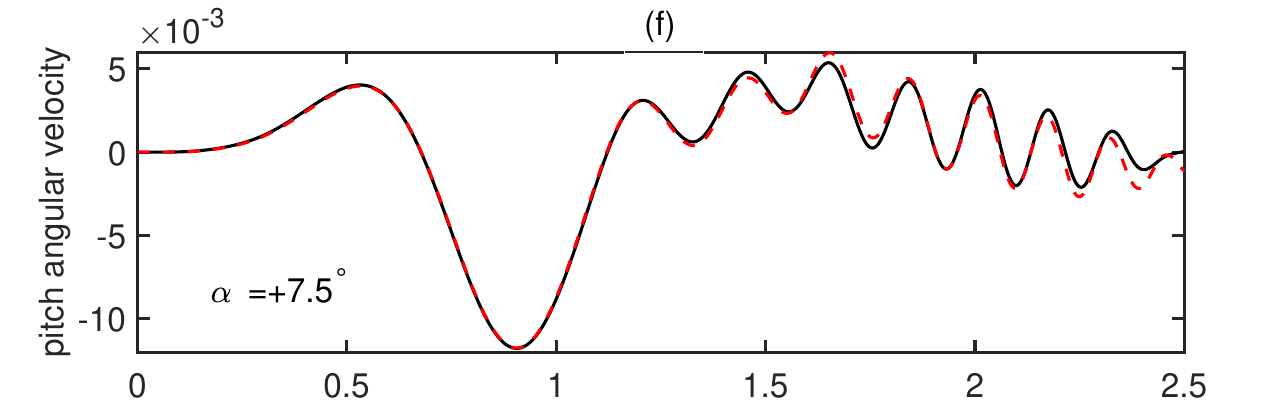}
		\label{fig:FlexVh_pitchrate_MoInd31}
	\end{subfigure}
	
	\caption{Time-domain responses of LPV-ROM and FOM at angle of attack $\alpha = -7.5, +7.5^{\circ}$. (a)-(b): Bending strains at wing root; (c)-(d): Longitudinal velocity; (e)-(f): Pitch rate.}
	\label{fig:FlexVh_strain_pitch_longvel_fixed}
\end{figure}

The time-domain responses of data-driven LPV-ROM and full-order linearized model are compared in Fig.~\ref{fig:FlexVh_strain_pitch_longvel_fixed}. The strain of wing root, longitudinal velocity, and pitch rate all match with an error less than $2\%$. Note that at $\alpha = -7.5, 7.5^{\circ}$, the aeroelastic model becomes unstable (see Fig. \ref{fig: FlexVh_root_loci_ROM}, one pair of eigenvalues have positive real parts), but the p-DMD can still render an accurate data-driven model. The frequency-domain response comparisons are omitted for the sake of space limit.

\subsection{Varying Scheduling Parameter (Angle of Attack)}
\label{verification_varying_AoA}

In this section, the capability to model aeroelasticity at varying angles of attack and non-equilibrium conditions is demonstrated. 

The nonlinear equation of motion Eq.~\eqref{eq: nl_ae_fd_eom} is numerically integrated in the time domain to obtain the vehicle's nonlinear transient behavior. The nominal flight speed is still \qty{22}{\meter\per\second} at the altitude of \qty{20000}{\meter}. The simulation starts with a static operating condition, where the body pitch angle is $\theta_B = \ang{2.14}$, elevator deflection is $\delta_e = \ang{2.57}$, and thrust is $T = \qty{51.07}{\newton}$. A chirp signal with a magnitude of $5^{\circ}$, sweep frequency $0.01$-$2$ Hz is applied to the elevator to excite the rigid-body pitch motions and aeroelastic dynamics. The simulation is conducted with a sampling time of $0.005$s, and the full-order nonlinear model is simulated within a $10$-second period. All the state and input data are collected. The scheduling parameter (angle of attack) is calculated in real-time by $\alpha  = - \arcsin(\beta_3/U_{\infty})$. 

\begin{figure}[h]
	\centering
	\includegraphics[trim={0in 3.5in 0in 3.5in},clip,width=0.75\textwidth]{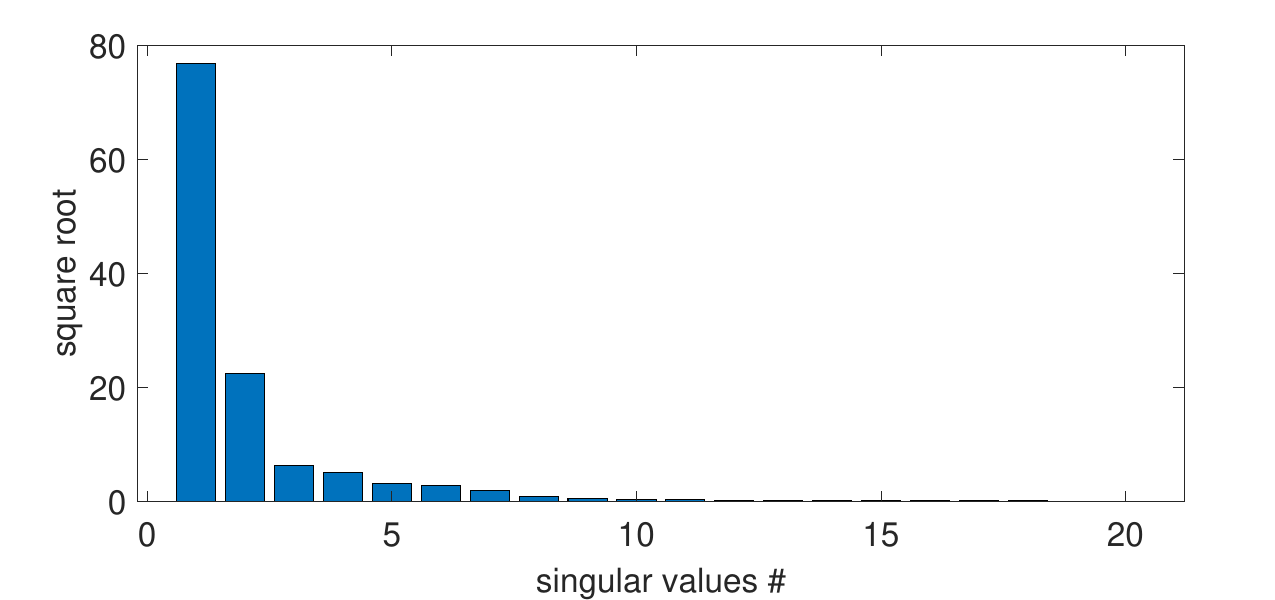}
	\caption{Square roots of 20 largest singular values  of the parametric Hankel matrix.}
	\label{fig:FlexVh_vary_singularValue_bar}
\end{figure}

The collected data snapshot is applied with the p-DMD to obtain the data-driven LPV-ROM. The 20 largest singular values of the parametric Hankel matrix are plotted in Fig. \ref{fig:FlexVh_vary_singularValue_bar}. The data-driven LPV-ROM is chosen to have the polynomial order of 4, and dimension of 12. 

\begin{figure}[ht]
	\centering
	\begin{subfigure}[b]{0.75\linewidth}
		\centering
		\vspace{-1em}
		\includegraphics[trim={0.0in 4.5in 0in 4.0in},clip,width=\textwidth]{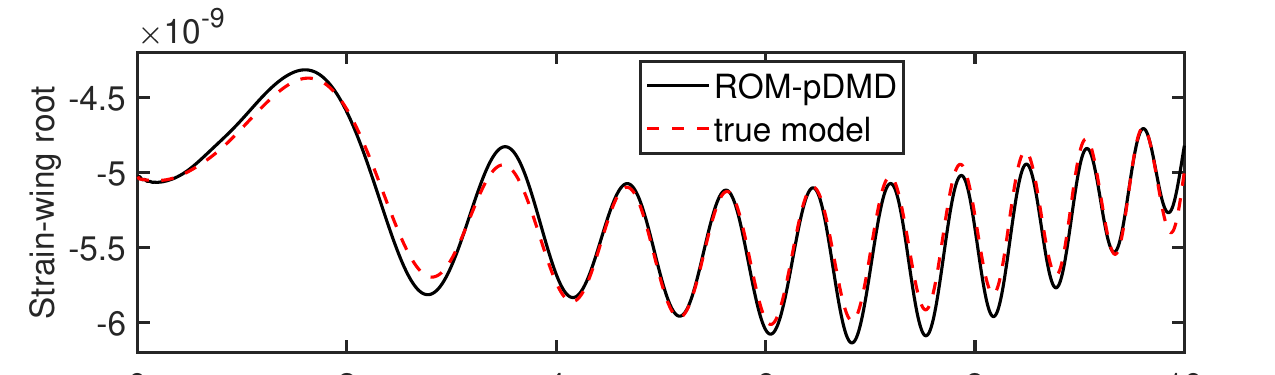}
		\label{fig:FlexVh_vary_strain_root_Morder12_test3rd}
	\end{subfigure}
	\begin{subfigure}[b]{0.75\linewidth}
		\centering
			\vspace{-1.85em}
		\includegraphics[trim={0.0in 4.5in 0in 4.0in},clip,width=\textwidth]{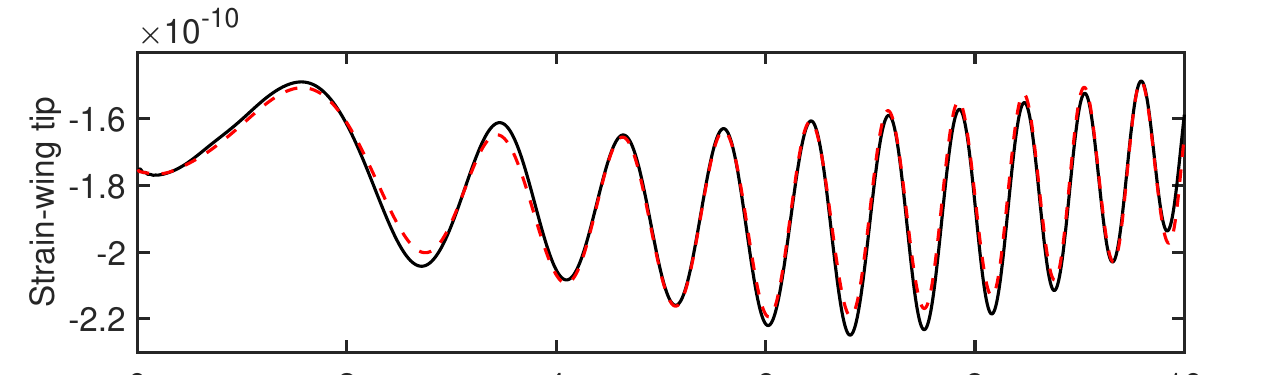}
		\label{fig:FlexVh_vary_strain_tip_Morder12_test3rd}
	\end{subfigure}

	\begin{subfigure}[b]{0.75\linewidth}
		\centering
				\vspace{-1.9em}
		\includegraphics[trim={0.0in 4.5in 0in 4.0in},clip,width=\textwidth]{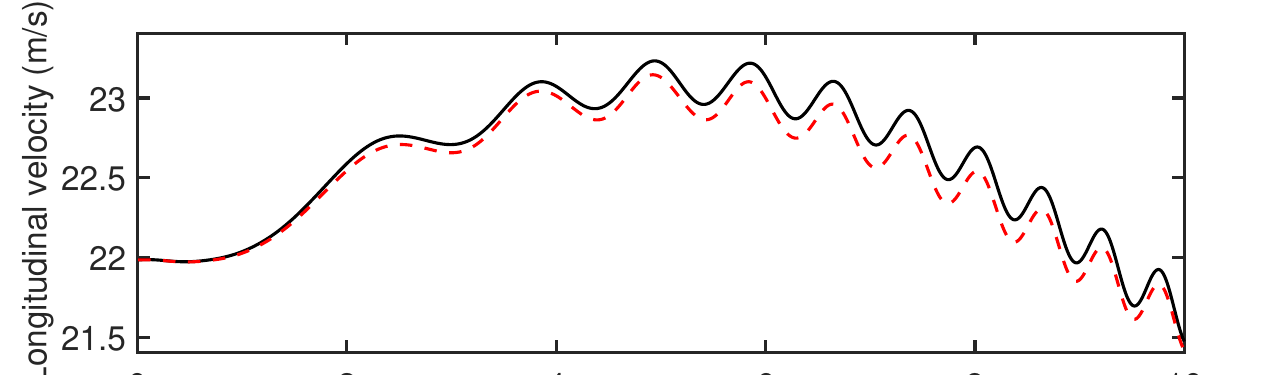}
		\label{fig:FlexVh_vary_long_vel_Morder12_test3rd}
	\end{subfigure}

	\begin{subfigure}[b]{0.75\linewidth}
		\centering
			\vspace{-1.8em}
		\includegraphics[trim={0.0in 4.5in 0in 4.0in},clip,width=\textwidth]{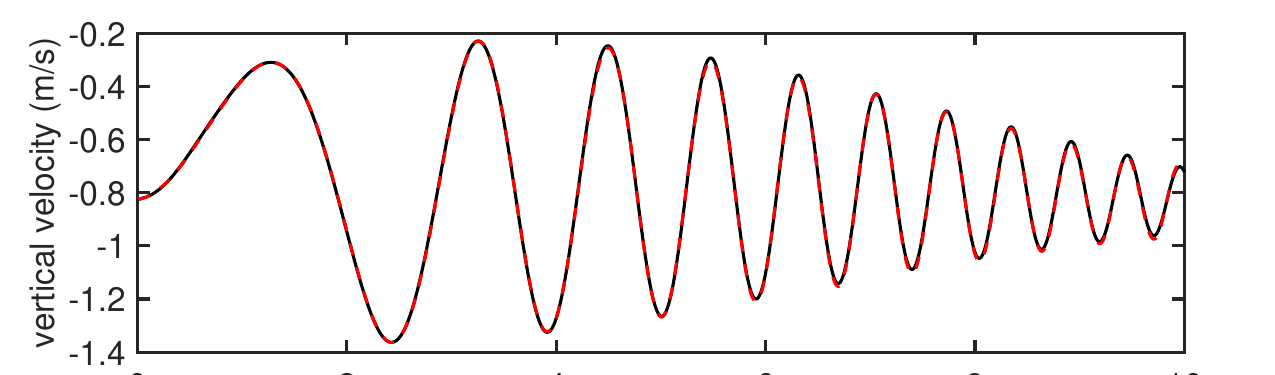}
		\label{fig:FlexVh_vary_vert_vel_Morder12_test3rd}
	\end{subfigure}

	\begin{subfigure}[b]{0.75\linewidth}
		\centering
				\vspace{-1.9em}
		\includegraphics[trim={0.0in 4.5in 0in 4.0in},clip,width=\textwidth]{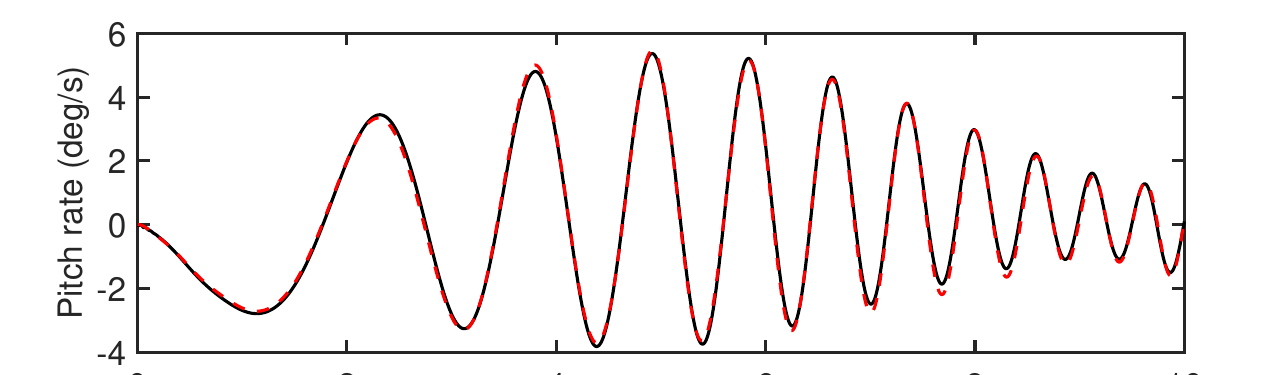}
		\label{fig:FlexVh_vary_PitchRate_Morder12_test3rd}
	\end{subfigure}

	\begin{subfigure}[b]{0.75\linewidth}
		\centering
				\vspace{-1.9em}
		\includegraphics[trim={0.0in 4.0in 0in 4.0in},clip,width=\textwidth]{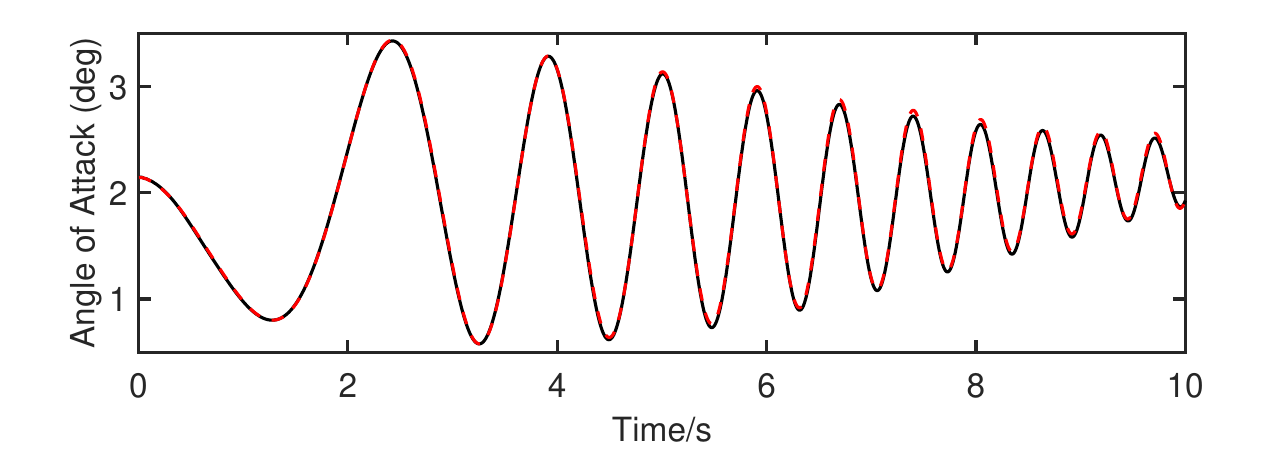}
		\label{fig:FlexVh_vary_AoA_Morder12_test3rd}
	\end{subfigure}
	\caption{Time-domain responses of LPV-ROM and nonlinear FOM at non-equilibrium flight condition.}
		\label{fig:FlexVh_varying}
\end{figure}

After that, the model accuracy of LPV-ROM is verified using other control inputs. Another chirp signal with a magnitude of $5^{\circ}$ and sweep frequency $0.1$-$2$ Hz is fed into the full-order nonlinear model and LPV-ROM, and their responses are compared. Fig.~\ref{fig:FlexVh_varying} shows the vehicle's time-domain elastic and rigid-body responses. Two models yield reasonably similar responses with a model error of $2.7\%$. From the results, it is worth emphasizing that the bending strains and rigid-body velocities start from a non-zero initial condition and remain at a non-equilibrium flight condition. This demonstrates that the proposed p-DMD can address the non-equilibrium conditions. In addition, the angle of attack calculated from the two models' responses matches accurately in Fig.~\ref{fig:FlexVh_varying}, which continuously varies between $0.5^{\circ}$-$3.5^{\circ}$. This study demonstrates that the proposed p-DMD can address the case of varying scheduling parameters. As stated in the novelties of this work, since no interpolation and linearization are used in the data-driven approach, the proposed method can derive parametric ROMs for flexible aircraft with large wing deformations and unsteady dynamics.

\section{Conclusion}\label{sec_conclusion}
This paper presented a data-driven approach for generating accurate linear parameter-varying reduced-order models (LPV-ROM) regarding the nonlinear aeroelasticity and flight dynamics of highly flexible aircraft. This method encoded the physical understanding of the nonlinear dynamics of highly flexible aircraft into a parametric Hankel matrix and performed parametric dynamic mode decomposition (p-DMD) to render LPV-ROMs. Both the fixed and varying scheduling parameter scenarios could be addressed by the p-DMD. The method was applied to and verified with a highly flexible cantilever wing and a complete free-flight aircraft. The local modeling accuracy with fixed scheduling parameters was verified by cross-validation with the linearized model. The global modeling accuracy was verified by comparing it with the nonlinear full-order model regarding the time-domain responses in transient flight conditions. This method is promising to use data from high-fidelity analysis to directly develop accurate models for flexible aircraft.

\bibliographystyle{ieeetr}
\bibliography{references}

%
%
%
\vfill

\end{document}